\documentclass[fleqn,usenatbib]{mnras}

\usepackage{graphicx}
\usepackage{amsmath}
\usepackage{amssymb}
\usepackage{newtxtext,newtxmath}

\usepackage{amsmath,amstext}
\usepackage{comment}
\usepackage[figure,figure*]{hypcap}
\usepackage{multirow}
\usepackage{amsmath}
\usepackage{nccmath}
\usepackage{rotating}
\usepackage{soul}
\usepackage{booktabs}
\setstcolor{red}
\usepackage{float}
\usepackage{longtable}
\usepackage{dcolumn}
\usepackage{orcidlink}
\usepackage{xspace}
\newcommand{\nodata}{...}

\newcommand{\orbitize}{\textit{Orbitize!}\xspace}

\defcitealias{Ortiz2017}{OLK17}
\defcitealias{dzib2010}{DLM10}
\defcitealias{loinard2008}{LTM08}
\defcitealias{loinard2007}{Loinard et al. 2007}
\defcitealias{ortiz2017b}{Ortiz-León et al 2017b}

\title[Dynamical mass of EC\,95]{\centering Dynamical mass of the Serpens intermediate-mass young stellar system EC\,95 with DYNAMO--VLBA}

\author[J. Ordóñez-Toro et al.]{%
\parbox{\textwidth}{
Jazmín Ordóñez-Toro$^{1}$\thanks{E-mail: n.ordonez@irya.unam.mx}\orcidlink{0000-0001-7776-498X}, 
Sergio A. Dzib$^{2}$\orcidlink{0000-0001-6010-6200}, 
Laurent Loinard$^{1,3,4}$\orcidlink{0000-0002-5635-3345}, 
Gisela Ortiz-León$^{5}$\orcidlink{0000-0002-2863-676X},
Marina A. Kounkel$^{6}$\orcidlink{0000-0002-5365-1267},
Phillip A. B. Galli$^{7}$\orcidlink{0000-0003-2271-9297}, 
Josep M. Masqué$^{8}$\orcidlink{0000-0002-1963-6848}, 
Luis H. Quiroga-Nu\~nez$^{9}$\orcidlink{0000-0002-9390-955X},
Sundar Srinivasan$^{1}$\orcidlink{0000-0002-2996-305X}, 
S.-N. X. Medina$^{2,10}$\orcidlink{0000-0003-2580-4796}, 
Luis F. Rodr\'{\i}guez$^{1}$\orcidlink{0000-0003-2737-5681},\\ 
}\\
$^{1}$Instituto de Radioastronom\'{\i}a y Astrof\'{\i}sica, Universidad Nacional Aut\'onoma de M\'exico, Apartado Postal 72-3, Morelia 58089, M\'exico\\
$^{2}$Max Planck Institut f\"ur Radioastronomie, Auf dem Hügel 69, D-53121 Bonn, Germany\\
$^{3}$Black Hole Initiative at Harvard University, 20 Garden Street, Cambridge, MA 02138, USA\\
$^{4}$David Rockefeller Center for Latin American Studies, Harvard University, 1730 Cambridge Street, Cambridge, MA 02138, USA\\
$^{5}$Instituto Nacional de Astrofísica, Óptica y Electrónica, Apartado Postal 51 y 216, 72000 Puebla, México\\
$^{6}$Department of Physics and Astronomy, University of North Florida, 1 UNF Dr., Jacksonville, FL, 32224\\
$^{7}$Instituto de Astronomia, Geofísica e Ciências Atmosféricas, Universidade de São Paulo, Rua do Matão, 1226, Cidade Universitária, 05508-090, São Paulo-SP, Brazil.\\
$^{8}$Departamento de Astronom\'{\i}a, Universidad de Guanajuato, Apartado Postal 144, 36000 Guanajuato, M\'exico\\
$^{9}$Department of Aerospace, Physics and Space Sciences, Florida Institute of Technology, 150 W University Blvd, Melbourne, 32901, FL, USA\\
$^{10}$German Aerospace Center, Scientiﬁc Information, 51147 Cologne, Germany\\
}
\date{}
\begin{document}
\maketitle
\begin{abstract}
We present dynamical mass measurements, obtained from multi-epoch Very Long Baseline Interferometry (VLBI) observations, for the young multiple stellar system EC\,95, located in the core of the Serpens star-forming region. Our dataset includes both archival data and more recent observations obtained as part of the \textit{Dynamical Masses of Young Stellar Multiple Systems with the VLBA project} (DYNAMO--VLBA), totalling 32 epochs over 12 years of observation. The observations span more than half of the EC\,95AB orbit, which has an estimated period of $21.6\pm0.1$ years, and yield masses of $2.15\pm0.10$ M$_\odot$ for EC\,95A and $2.00\pm0.12$ M$_\odot$ for EC\,95B. Furthermore, for the first time, we have estimated the mass of the third component (EC\,95C) using four available radio detections as well as an infrared observation. We find it to be 0.26 $^{+0.53}_{-0.46}$ M$_\odot$, suggesting that E\,95C is a low-mass T Tauri star. We conclude that EC\,95 is a hierarchical triple system comprised of a tight central binary proto-Herbig\,AeBe system and a lower mass companion on a wider orbit. These results contribute to improve our understanding of the stellar dynamics in this system and provide valuable insights into its triple nature.
\end{abstract}
\begin{keywords}
astrometry --- binaries--- stars: formation ---stars: kinematics and dynamics
\end{keywords}
\nopagebreak 
\section{Introduction}\label{intro}

A crucial parameter for understanding star formation processes is the mass of the newly formed stars. Stellar mass is the most fundamental parameter for stellar evolution, and its measurement during the early stages provides invaluable information on the amount of material available for star formation or the speed of molecular cloud collapse, among other key aspects. The most direct way to determine stellar masses is through analyzing the gravitational effects a star exerts on other objects in its vicinity; for example, a close companion \citep[e.g.,][]{duchene2013} or a disk \citep[e.g.,][]{tobin2012}. As they do not rely on stellar properties assumptions, dynamical masses are less biased and can be used to test evolutionary models. 

An interesting example of a multiple system in the Serpens region is EC\,95 \citep[][]{Eiroa1992},  
a young multiple stellar system embedded in the main core of Serpens at a distance of 436.0$\pm$9.2\,pc \citep{ortiz2018}. \citet{Preibisch1999} estimated the spectral type of EC\,95 as K2 based on infrared spectroscopic observations; he also derived an age of order $\sim 10^5$ years, a luminosity of $L=60^{+30}_{-20}$\,L$_\odot$, and a stellar radius of $R_\star=10.3\pm3.5$\,R$_\odot$. \citet{Preibisch1999} used available evolutionary models and noted that the location of EC\,95 in the HR diagram was consistent with a stellar mass of approximately 4\,M$_\odot$. He classified EC\,95 as a proto-Herbig AeBe star, pointing out that it takes $\sim10^6$\,years for a star of that mass to move roughly horizontally across the HR diagram from a nascent K spectral type to its final A-B type. 

Additionally, radio observations by \citet{Rodriguez1980}  and \citet{Eiroa2005} using the Very Large Array (VLA) detected EC\,95 as a strong, unresolved radio source, suggesting a non-thermal coronal origin for the radio emission. Subsequent radio observations with the Very Long Baseline Array (VLBA) by 
\citet[][ \citetalias{dzib2010} hereafter]{dzib2010} uncovered that the system consists of two compact components, with an approximate separation of 15 mas.
These observations were carried out over a time baseline of two years and showed that during that period, the motion of one of the stars was consistent with a linear motion. The motion of the other component had to be modeled with acceleration parameters, and an orbital period of 10–20 years was estimated. 
\citetalias{dzib2010} concluded that EC\,95 is a binary system where the primary is significantly more massive than the secondary. They named these stars EC\,95A and EC\,95B, associating the former with the proto-Herbig AeBe star and the latter with a low-mass companion. 
Additional VLBA observations of EC\,95  by \citet[][ \citetalias{Ortiz2017} hereafter]{Ortiz2017} increased the time baseline to eight years, offering a refined perspective on the dynamics of the system. The continued monitoring of the motion of both components showed that none was consistent with a linear motion and
provided a better dynamical mass determination by fully modeling the orbital motions of each components. This resulted in 
remarkably similar masses for the primary and secondary components, with values of 2.3$\pm$0.1\,M$_\odot$ and 2.0$\pm$0.2\,M$_\odot$, respectively. 

Adding complexity to this stellar arrangement, a third component, EC\,95C, was observed in the near-infrared observations conducted with the VLT in 2005 \citep{Duchene2007}. Recent VLBA observations obtained as part of the Gould Belt Distances Survey (GOBELINS) by \citetalias{Ortiz2017} also detected EC\,95C as a radio source located $\sim$145 mas northeast of the central binary system barycenter in September 2008 and $\sim$138 mas in January 2012. Consequently, EC\,95 emerges as a triple hierarchical system, where each component exhibits non-thermal radio emission.

\section{Observations, data reduction, and astrometric fitting procedures}

\subsection{Observations and data reduction}
Radio continuum observations of EC\,95 were carried out using the Very Long Baseline Array (VLBA) of the National Radio Astronomy Observatory (NRAO) as part of the DYNAMO-VLBA project BD215, at a wavelength $\lambda$ = 6.0 cm ($\nu$ = 5 GHz). The observational strategy consisted of six sessions separated from one another by about four months, starting in early 2018 and concluding in late 2019. The fourth observation, BD215E3 obtained in March 2019, was repeated as BD215E6 three days after the original attempt to account for an antenna misalignment -- BD215E3 was nevertheless usable and is included in the present analysis. As a result, EC\,95 was observed a total of seven times as part of the DYNAMO-VLBA project. Details regarding the observations and measured positions can be found in Table \ref{Ts1_full}. Each observation session lasted approximately three hours, with around 25 minutes allocated at the beginning and the end to observing a dozen quasars distributed across the sky. The remaining time was dedicated to the target source EC\,95, interspersed with scans of the main phase calibrator J1833+0115, located at $\sim0\rlap{.}^{\circ}79$ from EC\,95, and scans of the secondary phase calibrators J1832+0118, J1826+0149, and J1824+0119.

Data calibration followed standard VLBI procedures with phase referencing using the Astronomical Image Processing System (AIPS) software \citep{Greisen2003}.  These procedures have been detailed in previous works \citepalias{loinard2007,dzib2010,ortiz2017b} and include standard phase and amplitude calibrations as well as calibrations of the groups delays \citep{reid2004}, clock delays and tropospheric terms. After calibration, the visibility data were imaged with a pixel size of 50 $\mu$as and natural weighting (ROBUST = 5 in AIPS). The ﬂuxes and positions were measured using a two-dimensional Gaussian fitting procedure (JMFIT task in AIPS); the resulting rms noise levels in the final images ranged from 0.02 to 0.11 mJy beam$^{-1}$ (Table \ref{Ts1_full}).

Prior to the DYNAMO-VLBA project, EC\,95 was observed in multi-epoch observations as part of the VLBA projects BL155, BL160 (P.I.: L. Loinard), and BD155 (P.I.: S. Dzib) at a frequency of 8.4 GHz. These observations, totaling 16 epochs, were complemented by those from the GOBELINS project (code BL175; P.I.: L. Loinard), comprising an additional nine epochs that were observed at 4.9 GHz \citepalias{Ortiz2017}. In total, 32 VLBA observations of EC\,95 from 2007 to 2019 are considered here, they constitute the most comprehensive dataset for EC\,95 to date.

\begin{table*}
		\caption{EC\,95A and EC\,95B measured positions and flux densities for the 32 VLBA observations.}
		\label{Ts1_full}
	\footnotesize
	\begin{center}
 \setlength{\tabcolsep}{2.8pt}
 \renewcommand{\arraystretch}{1.05}
	\begin{tabular}{lcccccccccc} \hline\hline
        &        &          &\multicolumn{3}{c}{EC\,95A}&\multicolumn{3}{c}{EC\,95B} & \\ \cmidrule(lr{.75em}){4-6}\cmidrule(lr{.75em}){7-9}
Project  &Date UT&	Date	& $\alpha$(J2000.0) &  $\delta$ (J2000.0) &$S_\nu  \pm \sigma_{s_v}$ &$\alpha$(J2000.0) &  $\delta$ (J2000.0) &$S_\nu  \pm \sigma_{s_v}$ &$\sigma_{\rm noise}$	 \\  
name &(yyyy.mm.dd)& Julian Day& $18^{h}29^{m}$ [$^{s}$]&$1^{\circ}$12$'$ [$''$]&(mJy) &$18^{h}29^{m}$ [$^{s}$]&$1^{\circ}$12$'$ [$''$]&(mJy) &{\footnotesize ($\mu$Jy\,bm$^{-1}$)} \\ \hline
BL156  & 2007.12.22 & 2454457.32  &57.8909599(13) &46.107905(36)  & $~~ 1.79\pm0.10$ & \nodata  & \nodata &\nodata &50\\
BL160A   &2008.06.29 & 2454646.82 & 57.8909585(48)  & 46.107242(186)& $~~0.47\pm0.16$ &\nodata  & \nodata &\nodata  & 90\\
BL160B  & 2008.09.15 & 2454724.61 &57.8908095(8) & 46.105900(29) & $~~4.51\pm0.23$ &  \nodata & \nodata & \nodata &110\\
BL160C  & 2008.11.29 & 2454800.40 &57.8908841(22) & 46.104416(89) & $~~1.03\pm0.08$  &  57.8918664(18)&  46.110101(69) & $~~1.24\pm0.08$ &80\\
BL160D  & 2009.02.27 & 2454890.14 &57.8911210(39) & 46.103859(138) & $~~0.41\pm0.16$ & 57.8921732(5) & 46.106940(18) & $~~4.01\pm0.15$ &80\\
BL160E  & 2009.06.03 & 2454985.89 & 57.8910697(41) & 46.104177(240) &$~~1.57\pm0.24$ &\nodata  & \nodata &\nodata &110\\
BL160F  & 2009.08.31& 2455074.65 &  57.8909119(8)&  46.103134(32)& $~~3.11\pm0.14$ & \nodata  & \nodata &\nodata&70\\
BL160G  & 2009.12.05& 2455171.38 & \nodata & \nodata &\nodata  &        57.8922233(10)  &46.095333(41)  &$~~2.77\pm0.16$ &90\\
BL160H  & 2010.03.12& 2455268.12 &  57.8913181(40)&  46.100962(162)&$~~0.18\pm0.06$ &  57.8924800(20) &  46.092081(67) & $~~0.60\pm0.07$ &62\\
BL160I  & 2010.06.09& 2455356.88  & 57.8912856(36)&  46.101013(176)& $~~0.77\pm0.19$&  57.8924296(12) & 46.089683(54) & $~~1.72\pm0.16$ &74\\
BL160J  & 2010.09.03& 2455442.64 &  57.8911673(47)&  46.099877(202)&$~~0.32\pm0.10$   & \nodata & \nodata & \nodata &57\\
BD155A  & 2012.01.09 & 2455936.29  &57.8918555(61)&  46.091786(156)&$~~0.39\pm0.09$  &  57.8925187(46)  & 46.068868(150) &$~~0.56\pm0.10$ &49\\
BD155B  & 2012.01.10 & 2455937.29  &\nodata&  \nodata&\nodata  & 57.8925323(30)  & 46.068531(82) &$~~0.54\pm0.07$&41\\
BD155C  & 2013.08.18& 2456522.69 &  57.8924695(6)&  46.081657(23)&$~~1.37\pm0.05$ &   57.8923984(10)    & 46.053528(34) & $~~0.91\pm0.05$ &26\\
BD155D  & 2013.08.20& 2456524.68 &  57.8924686(42)&  46.081514(137)&$~~0.22\pm0.05$  & 57.8923894(23)  & 46.053868(110) &$~~0.38\pm0.05$ &22\\
BL175E2  & 2013.09.03& 2456538.71  & 57.8924432(69)&  46.081859(229)&$~~0.40\pm0.06$  & 57.8923300(15) & 46.054528(49) &$~~0.79\pm0.05$ &28\\
BL175G1  & 2014.03.03& 2456720.21  & 57.8929540(65)&  46.078949(167)&$~~0.24\pm0.05$ & 57.8926328(33)  & 46.049701(92) & $~~0.69\pm0.05$  &27\\
BL175CS  & 2014.10.13& 2456943.60  & 57.8929236(38)&  46.072521(108)& $~~0.34\pm0.04$ & 57.8923327(53)  & 46.043694(190) &$~~0.26\pm0.05$ &26\\
BL175FE  & 2015.03.02& 2457084.21 &  57.8933976(18)&  46.068654(71)& $~~1.04\pm0.06$ &\nodata & \nodata &\nodata &36\\
BL175GX  & 2015.10.07& 2457302.61  &57.8933947(77)  &46.063120(262) &$~~0.16\pm0.04$ &\nodata & \nodata &\nodata &25\\
BD155E   &2016.01.03& 2457391.31 &  57.8936486(1)&  46.060099(6)&$~~5.24\pm0.05$ &\nodata & \nodata & \nodata&25\\
BL175F8   &2016.04.28& 2457507.03  & 57.8938947(11)&  46.058656(36)&$~~1.17\pm0.04$  &  57.8926602(10) & 46.032795(36) & $~~1.18\pm0.04$ &24\\
BL175IM &2016.09.09& 2457640.67  & 57.8937768(10)&  46.054508(34)&$~~1.26\pm0.04$  & 57.8924047(7)  & 46.029802(25)&$~~1.67\pm0.04$ &19\\
BL175JF &2017.03.25  & 2457838.12 &57.8943146(30)&   46.048712(108)& $~~0.26\pm0.03$ & 57.8927517(13) & 46.025664(45) &$~~0.66\pm0.03$ &15\\
BL175KB &2017.09.16  & 2458012.65 &57.8942206(57)&  46.043325(181)&$~~0.40\pm0.03$  & 57.8924793(40) &  46.022171(130) &$~~0.58\pm 0.03$ &20\\
BD215 E0&2018.02.24  & 2458174.10 &57.8946426(141) & 46.039506(498)& $~~0.17\pm0.04$&57.8928053(13) & 46.018779(43) &$~~1.41\pm0.04$ &24\\
BD215 E1&2018.06.28  & 2458297.76  &57.8946522(59) &46.037134(231) & $~~0.13\pm0.03$  &57.8927006(15) & 46.018150(56) &$~~0.77\pm0.04$ &21\\
BD215 E2&2018.11.04  & 2458427.41  &57.8946539(21)  &46.031121(77) &$~~0.64\pm0.05$ &57.8926166(25)&46.014326(88)  &$~~0.56\pm0.05$  &26\\
BD215 E3&2019.03.04  & 2458547.09  &57.8950226(28)  &46.027755(115)& $~~0.52\pm0.06$ &57.8929021(55)   & 46.012754(240) & $~~0.14\pm0.04$ &36\\
BD215 E6&2019.03.07  & 2458550.07  &57.8950243(95)  &46.028162(322) &$~~0.14\pm0.05$ &57.8929306(46) & 46.012200(171) & $~~0.26\pm0.04$&28\\
BD215 E4&2019.07.12  & 2458676.73  &57.8949590(120)  &46.024994(356) & $~~0.16\pm0.04$&57.8928162(183)  & 46.010285(706) &$~~0.13\pm0.05$ &27\\
BD215 E5&2019.11.09  & 2458797.40  &57.8950192(92)  &46.020537(365) & $~~0.20\pm0.06$&57.8927570(9) & 46.007717(30) & $~~1.98\pm0.06$&45\\
\hline%
\end{tabular}  
 \end{center}
\end{table*}

\begin{table*}
\caption{EC\,95C measured positions and flux densities.}
\label{T_2}
\footnotesize
\begin{center}
 \setlength{\tabcolsep}{2.8pt}
 \renewcommand{\arraystretch}{1.05}
	\begin{tabular}{lccccccc} \hline\hline
\multicolumn{7}{c}{EC\,95C}\\
Project  &Date UT&	Date	& $\alpha$(J2000.0) &  $\delta$ (J2000.0) &$S_\nu  \pm \sigma_{s_v}$  &$\rho$\textsuperscript{a} &  PA\textsuperscript{a}  \\  
name &(yyyy.mm.dd)& Julian Day& $18^{h}29^{m}$ [$^{s}$]&$1^{\circ}$12$'$ [$''$]&(mJy)  & (mas) & ($^\circ$)\\ \hline \hline%
{VLT obs.}\textsuperscript{b} & 2005.05.22 & 2453512.85 & \nodata & \nodata &\nodata &152$\pm$1  &  47.2$\pm$0.5   \\\hline\hline
BL160B  & 2008.09.15& 2454724.61 & 57.8985675(3)& 46.205651(108)& $~~0.86\pm0.19$ &145.93$\pm$0.08&48.87$\pm$0.04\\
BD155A  & 2012.01.09 & 2455936.29  &57.8994536(6)&  46.166823(19)& $~~0.56\pm0.10$  & 138.81$\pm$0.08&51.64$\pm$0.04\\
BL175KB& 2017.09.16 & 2458012.65  &57.9004380(10)& 46.103750(30) & $~~2.28\pm0.03$  & 127.24$\pm$0.08&56.64$\pm$0.05\\
BD215 E1&2018.06.28  & 2458297.76  &57.9007544(136) &46.096234(606) & $~~0.30\pm0.03$ & 126.04$\pm$0.38&57.25$\pm$0.25\\
\hline
\end{tabular}  
 \end{center}
\begin{center}
        \footnotesize
        \textsuperscript{a} The separation ($\rho$) and position angle (PA; North through East) are taken with respect to the barycenter (VLBA) or the photocenter (IR) of the close binary EC\,95AB. 
        \textsuperscript{b} Reference: \citet{Duchene2007}. 
    \end{center}
\end{table*}

\subsection{Astrometric fitting procedures}

The motion of binary stars on the sky is determined by the combination of their trigonometric parallax, $\pi$, the uniform proper motion of their center of mass in right ascension and declination, denoted as $\mu_\alpha$ and $ \mu_\delta$ respectively, and their orbital motion:

\begin{eqnarray}
\alpha(t) & = & \alpha_0 + \mu_\alpha t + \pi f_\alpha(t) + a Q_\alpha(t),\label{eqn:pm1}\\
\delta(t) & = & \delta_0 + \mu_\delta t + \pi f_\delta(t) + a Q_\delta(t)\label{eqn:pm2}.
\end{eqnarray}

\noindent
Here, $f_\alpha(t)$ and $f_\delta(t)$ are the projections of the parallax ellipse in right ascension and declination, respectively, and $Q_\alpha(t)$ and $Q_\delta(t)$ describe the components of the orbital motion. The formulae are given explicitly in \citet{Ordonez-Toro}; they depend explicitly on the usual orbital elements: the semi-major axis, $a$, the orbital period, $P$, the eccentricity, $e$, the time of periastron passage, $T_0$, the inclination, $i$, the angle of the line of node, $\Omega$, and the angle from the ascending node to periastron, $\omega$. To derive the astrometric and orbital parameters of the EC\,95 system, we used the MPFIT least squares method implemented in \citet{kounkel2017} which fits the measured positions of each star with the equations of motion given above. It is important to emphasize that this method can use VLBI positions even when only one of the two stars is detected since it fits the positions of the individual stars -- in other words, it determines separately the semi-major axes $a_1$ and $a_2$ of the orbital paths of each star around the center of mass of the system, not just the semi-major axis $a$ of their relative motion. By the same token, it provides the mass of the individual stars (not just the total mass of the system); this is illustrated more clearly in the bottom panel of Figure \ref{EC95or}.

To independently confirm the orbital parameters, we also used the \orbitize\ package which combines the \textit{Orbits For The Impatient} (OFTI) methods with a \textit{Markov Chain Monte Carlo} (MCMC) fitting scheme \citep{orbitize}. Unlike the MPFIT algorithm mentioned above, \orbitize\ works with relative positions. As a consequence, the distance needs to be entered as a fixed parameter (in practice, we use the value obtained from the MPFIT method) and  \orbitize only returns the total mass of the system (not the individual masses). On the other hand, \orbitize treats the errors in a more rigorous way and has been shown to work well on VLBA observations \citep{Ordonez-Toro}.

\section{Results}
\begin{figure*}
\includegraphics[scale=0.58,trim=0 0 0 0,clip]{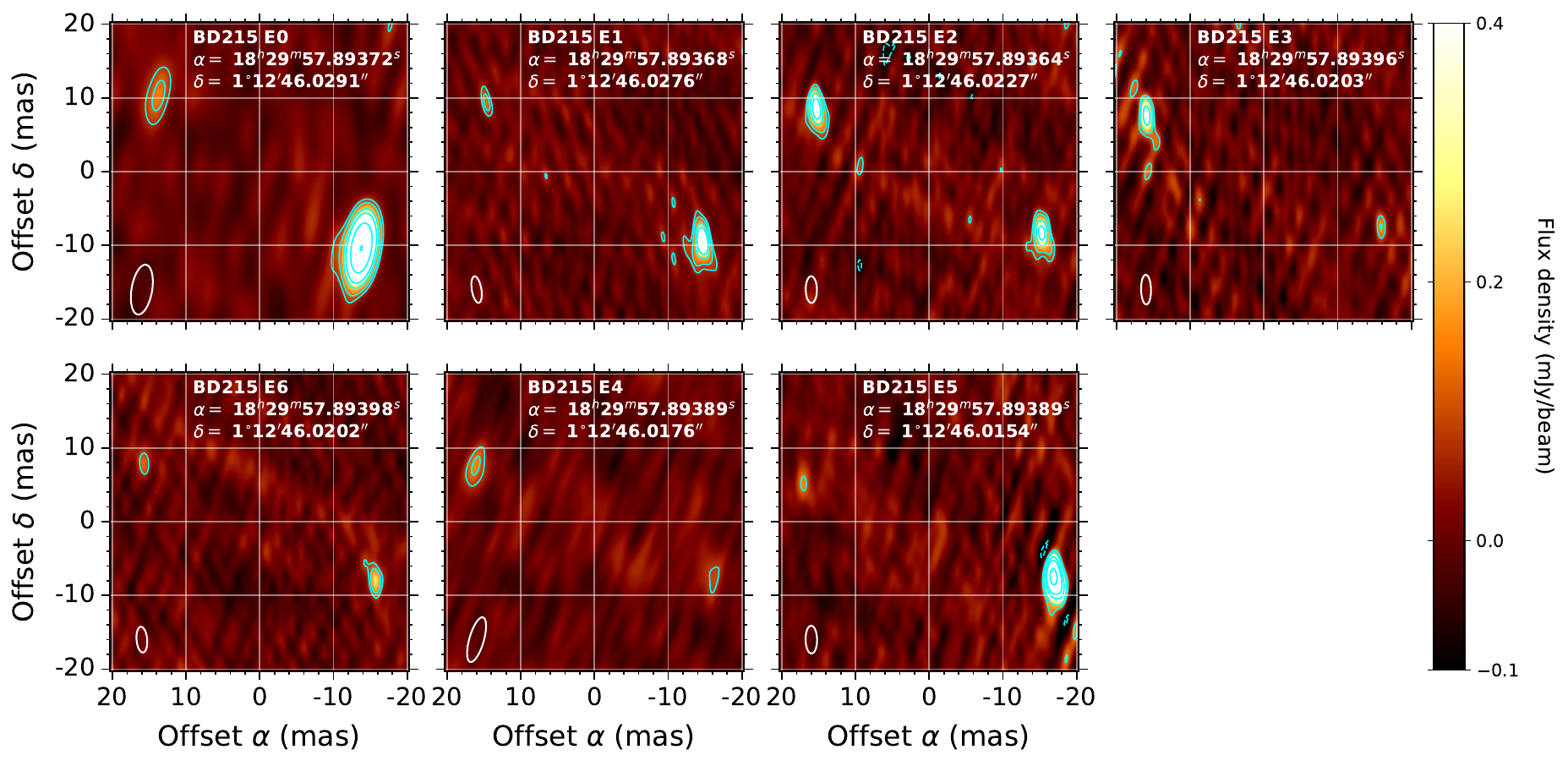}
\label{Ec95_epochs}
\caption{VLBA radio images of EC\,95A and EC\,95B at 4.9 GHz corresponding to each epoch observed during the DYNAMO-VLBA project. Intensity background images are clipped to intensities between --0.1 to 0.4 mJy beam$^{-1}$. Contour levels are at --3, 3, 6, 10, 15, 30, and 60 times the noise levels of the images, listed in Table~\ref{Ts1_full}. Images are centered at the mid-point between EC\,95A and EC\,95B at the coordinates indicated in the top-left corner of each plot. The white ellipse at the bottom- right of each plot represents the synthesized beam.}
\end{figure*}

\begin{figure}
\includegraphics[width=0.48\textwidth]{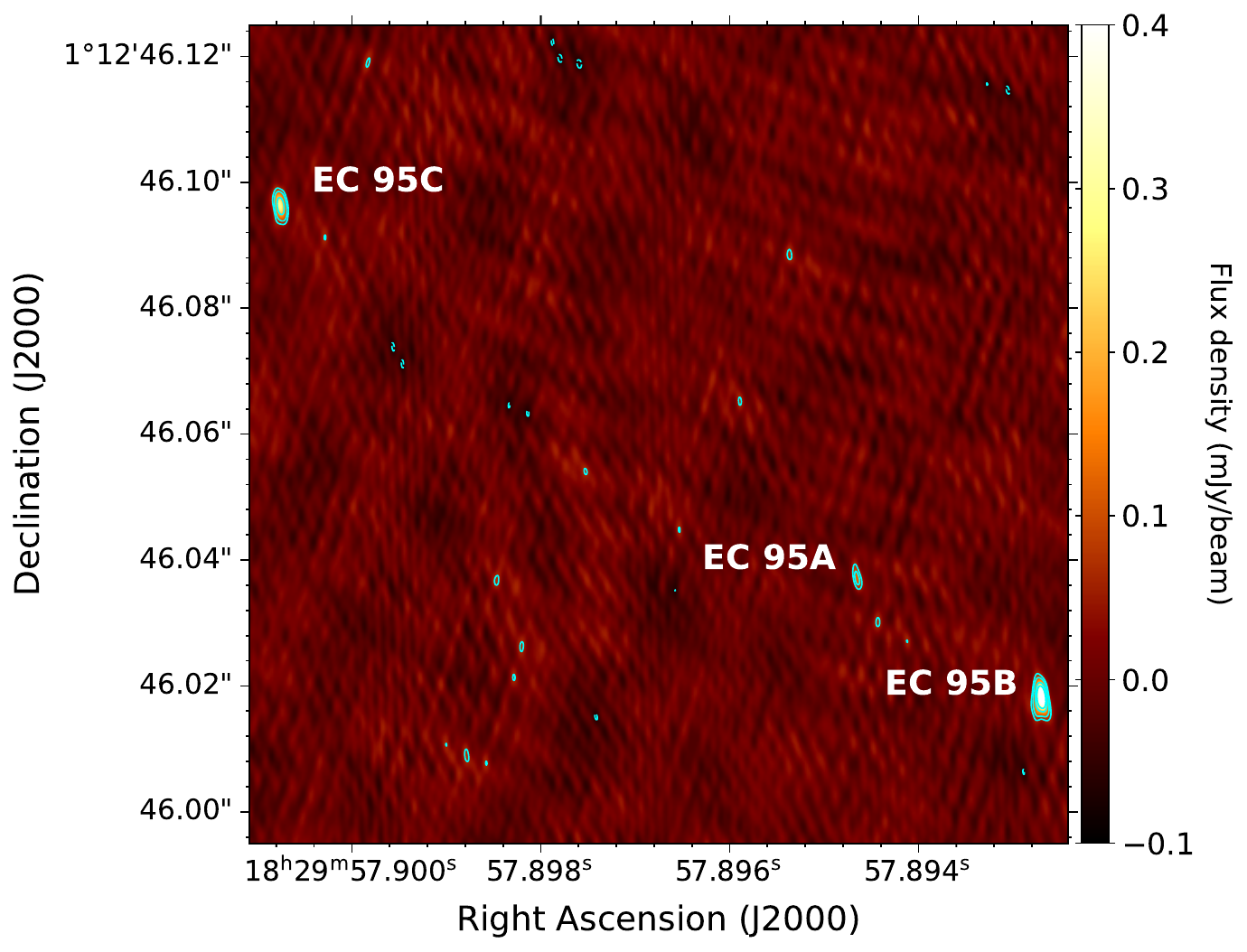}
\caption{VLBA radio image of EC\,95 at 4.9 GHz for epoch BD215E1, showcasing the detection of all three components (EC\,95A, EC\,95B, and EC\,95C) in the system. The contour levels are as in Fig.~\ref{Ec95_epochs}.}\label{Ec95_system}
\end{figure}
Components of EC\,95 were conclusively detected in all seven recent VLBA observations, as illustrated in Figure~\ref{Ec95_epochs}. The central binary components EC\,95A and EC\,95B were consistently identified at all epochs, with EC\,95A exhibiting an average flux density of 0.3 mJy, while EC\,95B displayed a mean flux density of 0.7 mJy during the DYNAMO-VLBA epochs. Table~\ref{Ts1_full} includes both the observations from the DYNAMO-VLBA project and those previously reported by \citetalias{dzib2010} and \citetalias{Ortiz2017}. The positions and flux densities for the latter were directly taken from these reports, which used the same source parameters extraction as in the present work. In total, EC\,95A was detected in 30 epochs, while the secondary component, EC\,95B, was detected in 23 epochs. The precise positions and flux densities for both EC\,95A and EC\,95B are concisely presented in Table~\ref{Ts1_full}.

A short disclaimer is in order here. The first resolved detections of EC\,95 were reported in \citetalias{dzib2010}. In that work, one of the two stars in the system showed a linear and uniform motion while the trajectory of the other was clearly curved. \citetalias{dzib2010} naturally concluded that the former was the more massive member of the system (the primary) while the latter was the less massive secondary, and called these two stars EC\,95A and EC\,95B, respectively. \citetalias{Ortiz2017} reported additional VLBI resolved observations and found that both stellar trajectories were curved indicating (i) similar masses and (ii) that the more massive member was actually the star that \citetalias{dzib2010} had identified as secondary -- the observations reported by \citetalias{dzib2010} were obtained during a time interval when the motion of the primary was, fortuitously, roughly linear and uniform. Thus, \citetalias{Ortiz2017} correctly identified the primary and the secondary, but called the primary EC\,95B and the secondary EC\,95A, for consistency with the nomenclature used by \citetalias{dzib2010}. Here, for consistency with common practice, we call the primary EC\,95A and the secondary EC\,95B. This implies that the names are reversed from the articles by \citetalias{dzib2010} and \citetalias{Ortiz2017}.

As previously mentioned, EC\,95 is a hierarchical triple system. Our observations from the DYNAMO-VLBA project resulted in one detection of the third component, complemented by three prior detections  \citep[\citetalias{Ortiz2017},][]{ortiz2018}. These data, together with an infrared position reported by \citet{Duchene2007}, allow us to carry out an initial analysis of the astrometry of this component, contributing to a more comprehensive understanding of the EC\,95 system as a whole. The measured properties of EC\,95C are provided in Table~\ref{T_2}. Figure~\ref{Ec95_system} shows the VLBA image corresponding to project BD215E1 where all three components in EC\,95 are detected.

\subsection{The masses of the close binary system EC\,95AB}

\begin{figure}
\includegraphics[scale=0.32]{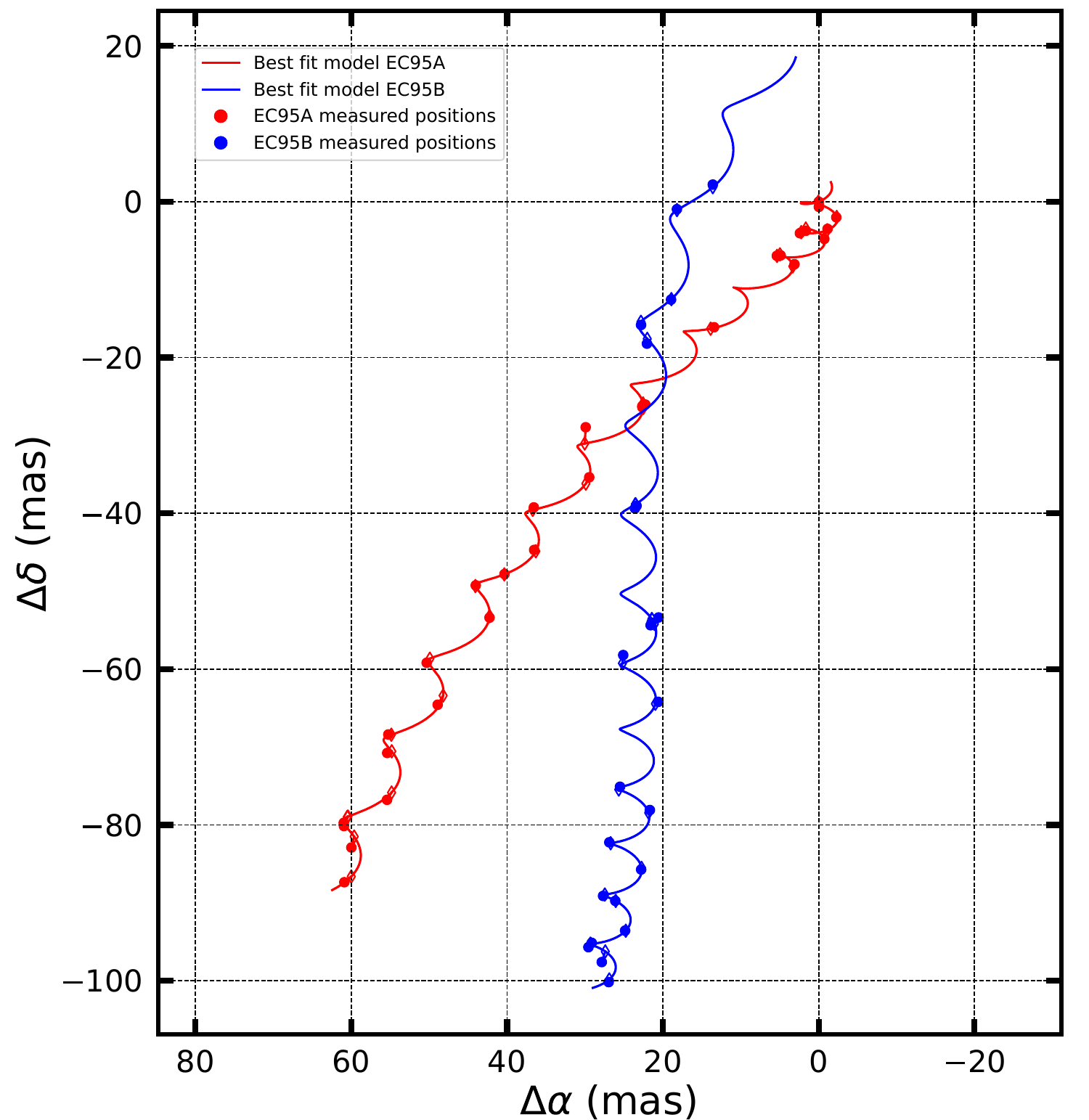}\\
\includegraphics[scale=0.32]{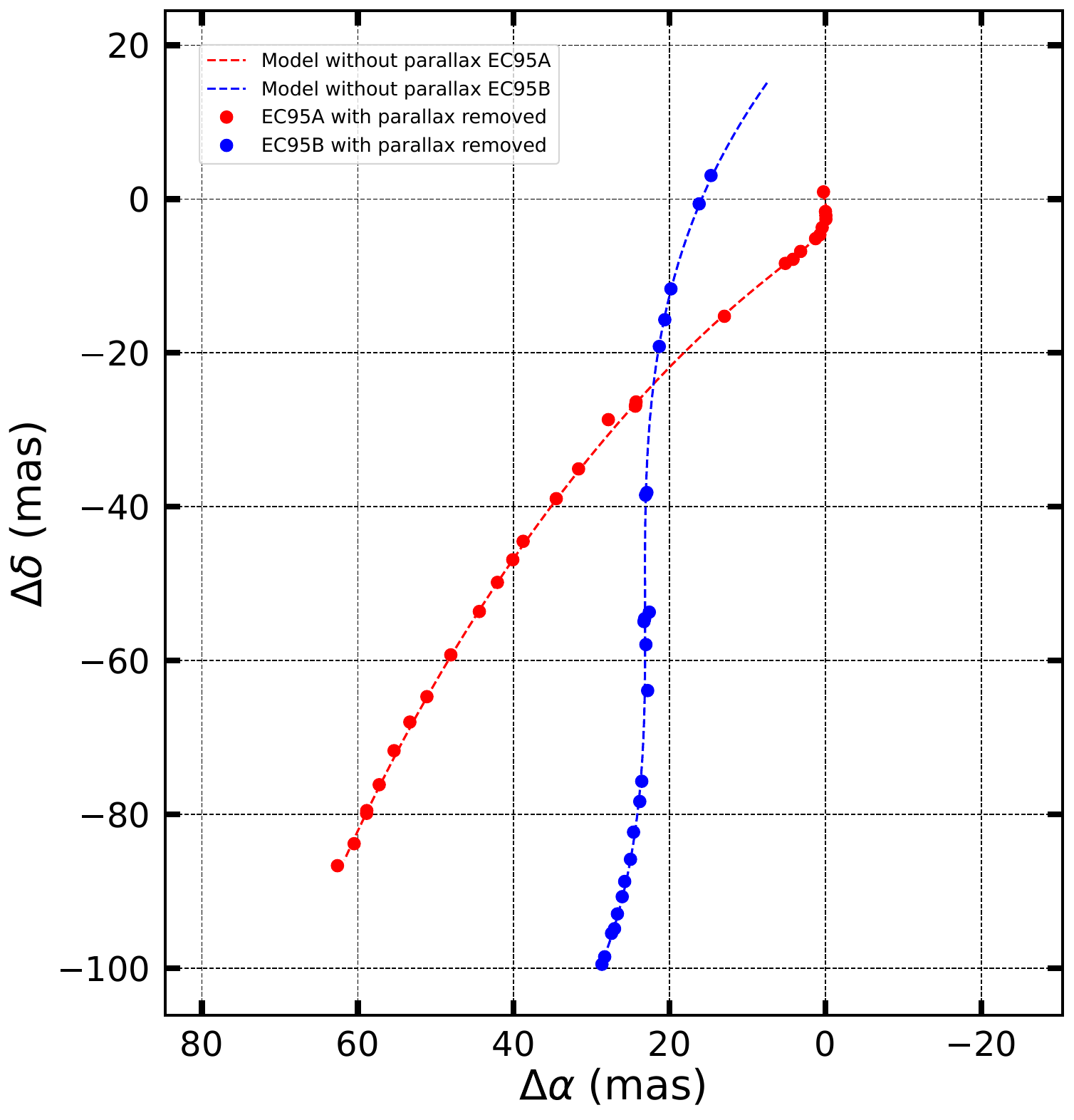}\\
\label{EC95sky}
\caption{Measured positions of EC\,95A (red dots) and EC\,95B (blue dots) shown as offsets from the position of EC\,95A in the first detected epoch in VLBA observations (2007 December 22, see Table~\ref{Ts1_full}). 
Top panel: The blue and red curves show the stellar motion best fits, described in the text, to the positions of EC\,95A and EC\,95B, respectively.  Bottom panel: The measured positions and best fit after removing the parallax signature.}
\end{figure}

\begin{figure}
\includegraphics[width=0.5\textwidth]{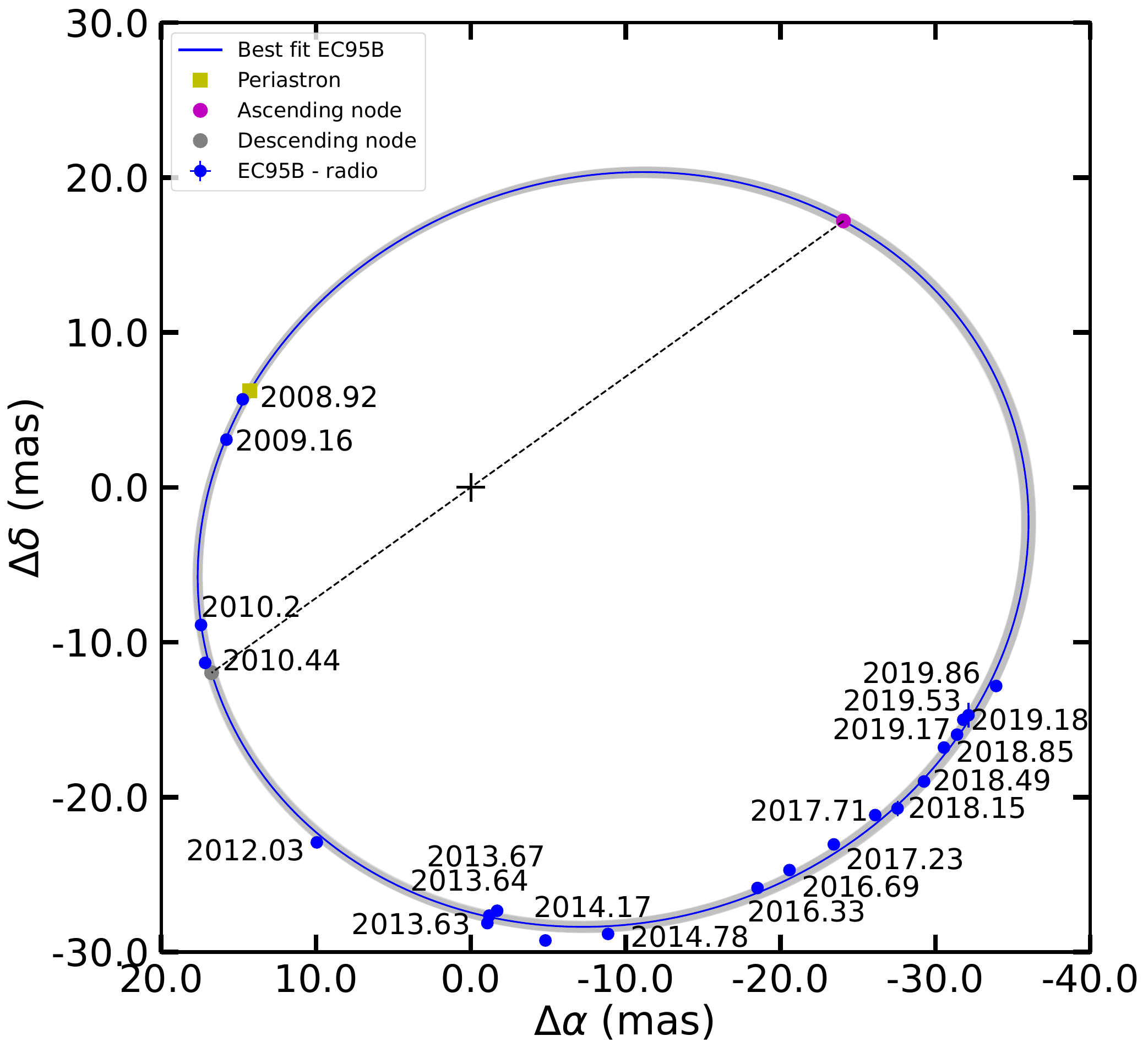}\\
\includegraphics[width=0.5\textwidth]{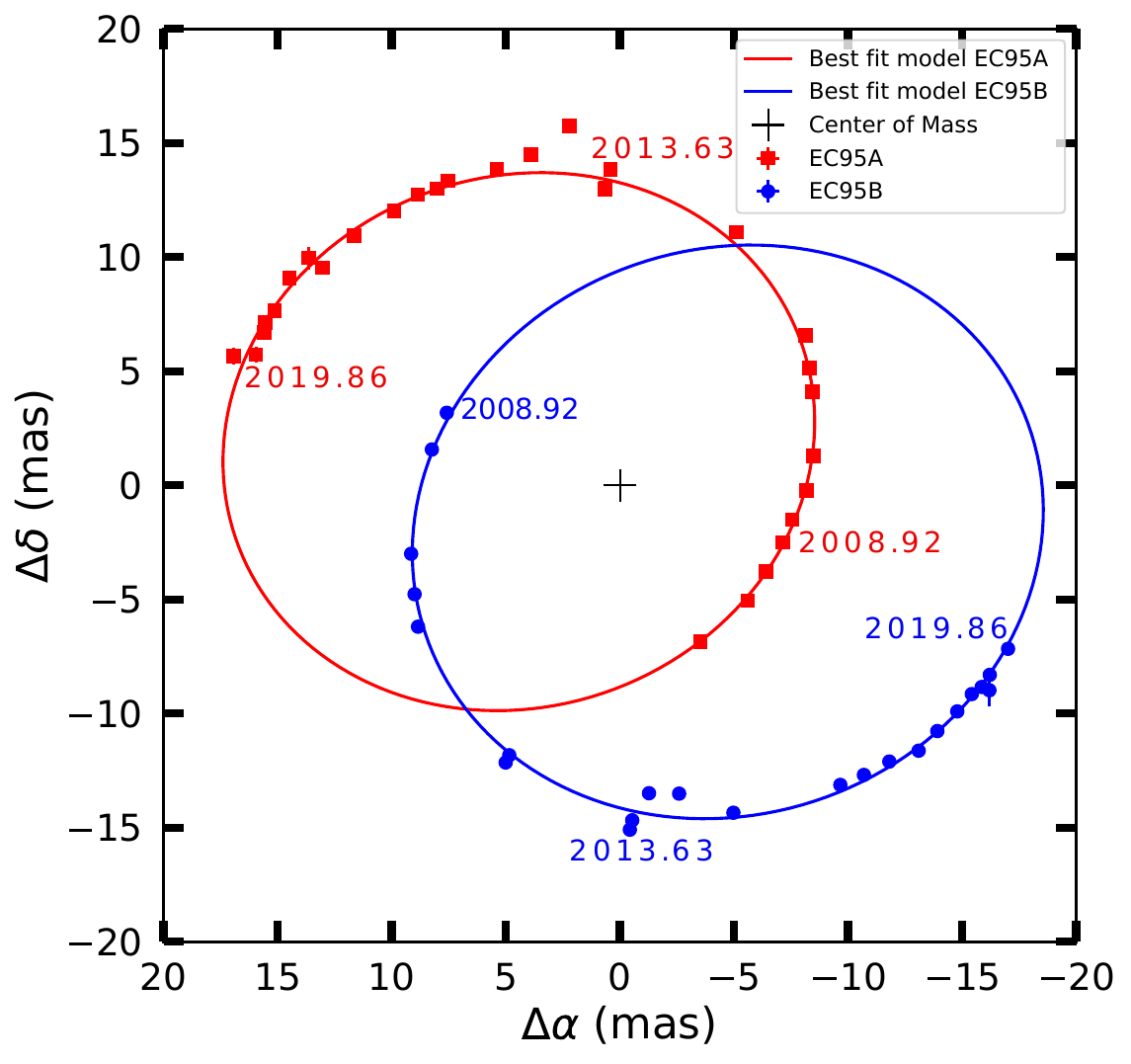}\\
\label{EC95or}
\caption{Top panel: Relative positions and orbital fit model for EC\,95AB. The blue dots indicate the relative positions of EC\,95B  with respect to EC\,95A, and the error bars consider the position errors of both components which are added in quadrature. The dashed black line traces the line of nodes from the model, and the black cross indicates the position of the primary source. Bottom panel: Stellar orbits for EC\,95AB are shown relative to the center of mass (black cross). }
\end{figure}

Starting from all the measured positions of EC\,95A and EC\,95B, we conducted an astrometric fitting using the MPFIT method. The results of the best fit are presented in Table~\ref{T_A0}, while the visualizations of the total sky motion and orbital motion are shown in Figures~\ref{EC95sky} and \ref{EC95or}, respectively. The best fit yields a mass of $2.148\pm0.097$ M$_\odot$ for the primary component and $1.998\pm0.116$ M$_\odot$ for the secondary, resulting in a total mass of $4.146\pm0.212$ M$_\odot$. We note that the observations span a total of 11.88 years, more than half the orbital period of $21.6\pm0.1$ years.

For the analysis with \orbitize, we used 23 observations in which both components were detected simultaneously. The position offsets and position angles of the components were employed as inputs for \orbitize Priors for the orbital elements were defined, and MCMC exploration was conducted with 10,000 walkers and 10,000 orbits. 
The results are presented in Table \ref{mcmc}; additionally, the corner plot showing the distribution posteriors is included in Figure \ref{corner_plot} of Appendix \ref{app:1}. Figure \ref{plot_mcmc} illustrates 2,000 orbits of the allowed orbital configurations, derived from the MCMC analysis. The resulting total mass of the close binary system is 4.52 $^{+0.25}_{-0.23}$ M$_\odot$, consistent within $1\sigma$ with the value obtained using MPFIT and with similar error bars.


\begin{table}
	\renewcommand{\arraystretch}{1.1}
		\caption{Best-fit model parameters for the close binary system EC\,95AB using MPFIT.}
		\label{T_A0}
	\centering
	\small
	\begin{tabular}{cccc} \hline \hline
		&Parameter& Value & Units 	  \\ 
		\hline
		\multicolumn{2}{l}{Astrometric parameters} &&\\
		& $\alpha_{2016.0, {\rm centre}}$ & 18:29:57.8931008(53)& hh:mm:ss\\
		& $\delta_{2016.0, {\rm centre}}$ & 1:12:46.048129(117) & $^\circ:\,':\,''$\\
		&$\mu_{\alpha}$  & $3.54\pm0.02$ & mas yr$^{-1}$\\	
		&$\mu_{\delta}$  & $-8.42\pm0.02$ & mas yr$^{-1}$\\
		&$\pi$           & $ 2.30\pm0.04$ & mas\\
            &d & $435.71\pm7.43$ & pc \\
		\multicolumn{2}{l}{Orbital parameters}&&\\
		&$a_1$ & $6.00\pm0.15$ & AU \\
		&$a_2$ & $6.46\pm0.09$ & AU \\
            &$a=a_1+a_2$ & $12.46\pm0.18$ & AU \\
	    &$P$ & $21.61\pm0.11$ & years\\
	    &$T_0$ & $2454779.43\pm33.00$ & Julian date \\
	    &$e$ & $0.391\pm0.003$ & \\
	    &$\Omega$ & $305.56\pm1.13$ & degrees \\
	    &$i$ & $30.44\pm0.26$ & degrees \\
	    &$\omega$ & $117.27\pm0.51$ &degrees \\
	    \multicolumn{2}{l}{Dynamical masses}&&\\
	    &$M_{A+B}$ & $4.146\pm0.212$ & M$_\odot$ \\
	    &$M_A$ & $2.148\pm0.097$ & M$_\odot$ \\
	    &$M_B$ & $1.998\pm0.116$ & M$_\odot$ \\ 
	    \hline
	\end{tabular}  
\end{table} 

\begin{table}
	\renewcommand{\arraystretch}{1.1}
\caption{Best-fit model parameters for the close binary system EC\,95AB using \orbitize The values of $\omega$ and $\Omega$ refer to EC\,95B.}
	\label{mcmc}
	\centering
	\small
	\begin{tabular}{cccc} \hline \hline
		& Parameter & Value & Units \\ 
		\hline
		\multicolumn{2}{l}{Orbital Parameters}&&\\
& $a$ & 13.51 $^{+0.26}_{-0.24}$ & AU \\
& $e$ & 0.393 $^{+0.002}_{-0.002}$ & \\
& $\Omega$ & 317.21 $\pm 0.82$& degrees\\
& $i$ & 35.44 $^{+0.34}_{-0.33}$ & degrees \\
& $\omega$ & 115.91 $\pm 0.38$& degrees\\
& $T_0$ & 2458862.0$\pm 0.25$& Julian date\\
& $P$ &23.37$\pm 0.22$& years\\
& $M_{A+B}$ & 4.52 $^{+0.25}_{-0.23}$ & $M_{\odot}$\\
	      
	    \hline
	\end{tabular}  
	
\end{table}

\newpage

For many orbital parameters, the values obtained from MPFIT and \orbitize are consistent within $2\sigma$. For others, however, the discrepancy is larger -- it is, for instance, $5\sigma$ for the orbital period. In part, these discrepancies are likely caused by the fact that different datasets are used for the two fits since only observations where both components are detected are used for \orbitize It is likely, however, that the observed discrepancies also partly reflect the fact that the orbital fits are not yet fully constrained by the data. Additional observations over the next decade will enable us to complete the monitoring of a full orbital period and ought to remove these inconsistencies.

\begin{figure*}
\includegraphics[scale=0.55]{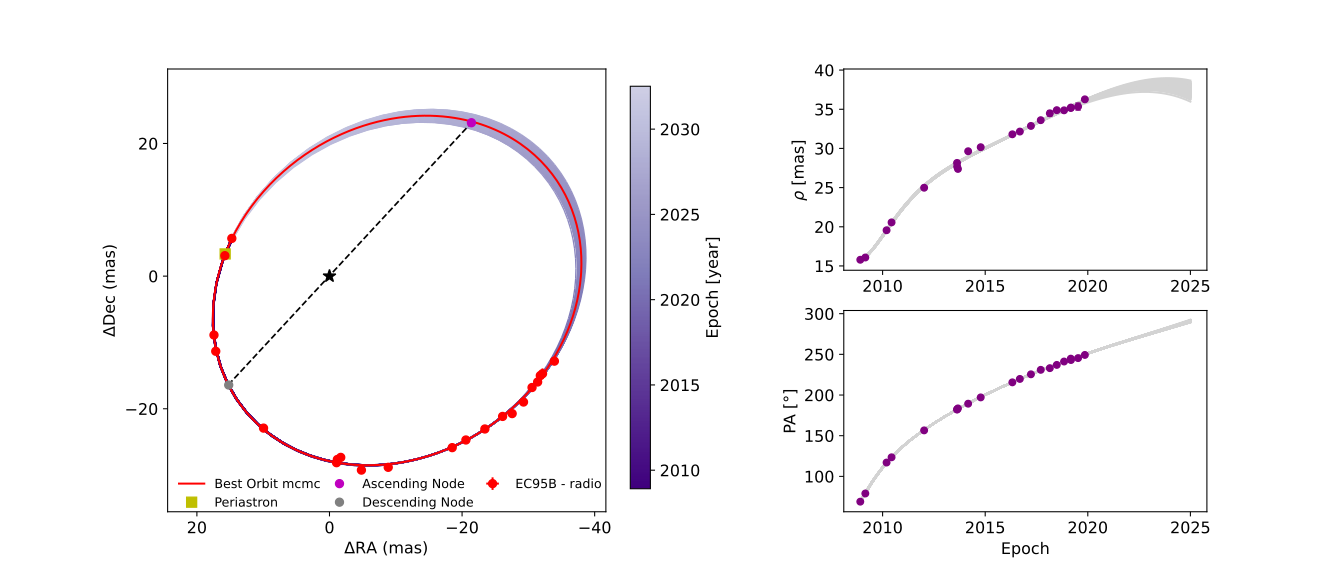}\\
\label{plot_mcmc}
\caption{Allowed orbital configurations for the close binary EC\,95AB derived from the MCMC analysis conducted using the \orbitize package. The red orbit corresponds to the best-fit orbit from the MCMC analysis. The black marker (star) designates the position of the primary component. The color bar indicates time measured from the first observation. The upper right panel depicts the evolution of the angular separation over time, while the lower right panel illustrates the variation in the position angle throughout the orbits. }
\end{figure*}

\subsection{The mass of the third component in the EC\,95 system}

Given the hierarchical triple nature of the EC\,95 system, its complexity can be simplified by modeling the central binary (EC\,95AB) as a single mass ($M_{A+B} = M_A+M_B$) located at the center of mass of the system. Thanks to the astrometric MPFIT results, we can accurately estimate the position of the center of mass at each epoch, and calculate the relative position of the third component, EC\,95C, for each VLBA detection. The corresponding separations and position angles are given in Table \ref{T_2}. Our VLBA detections span a period of approximately 10 years. Given the presumed long period of this component, estimated at approximately 260 years by \citetalias{Ortiz2017}, we complemented these observations with a near-infrared (NIR) detection obtained with the Very Large Telescope (VLT) in May, 2005, and reported by \citet{Duchene2007}. These observations do not resolve the tight central binary, so the position of EC\,95C in this case is referenced to the photocenter of the EC\,95AB system. This likely differs slightly from the center of mass, but the difference must be fairly small since the mass ratio of EC\,95AB is close to unity. Inclusion of this additional data point increases our time coverage slightly, to about 13 years. 

\begin{figure*}
\includegraphics[width=\textwidth]{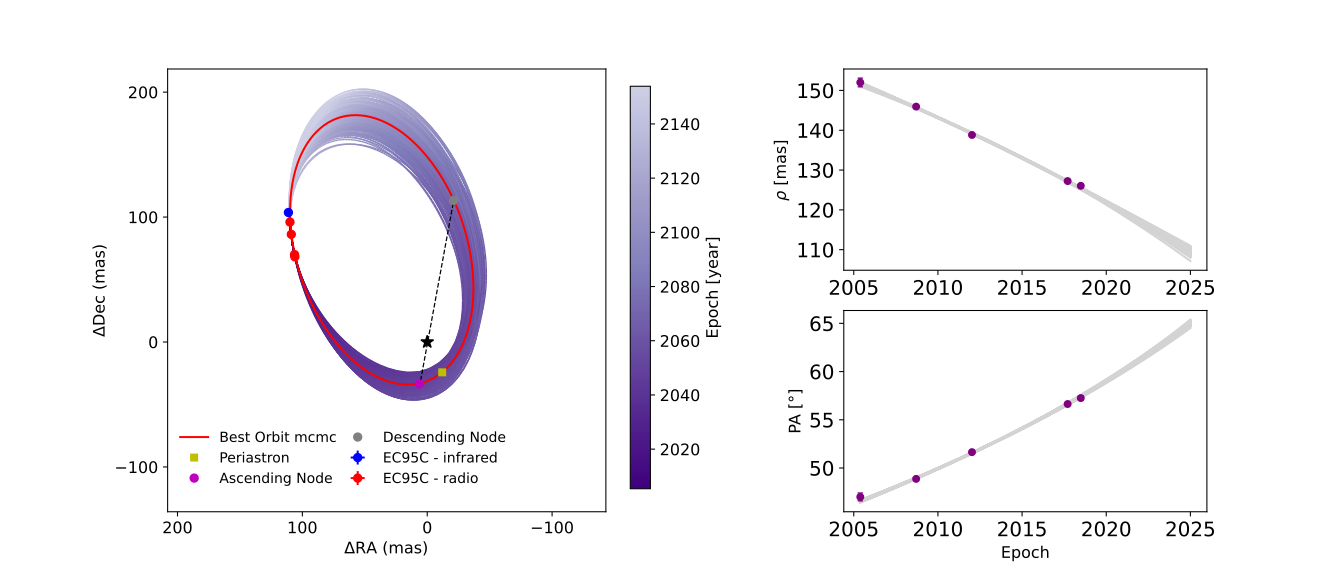}\\
\label{Orbits_ec95c}
\caption{Allowed orbital configurations for component EC\,95C, derived from MCMC analysis using \orbitize The blue point corresponds to the near-infrared (NIR) detection obtained at the VLT in 2005 \citep{Duchene2007}, while the red points represent radio detections detailed in Table \ref{T_2}. The red orbit corresponds to the best-fit orbit from the MCMC analysis. The color bar indicates time measured from the first observation. The upper right panel shows angular separation, and the lower right panel displays position angle variation during orbit.}
\end{figure*}

We applied the MCMC method implemented in the \orbitize package to these five detections (4 VLBA + 1 VLT). This tool is particularly useful in situations with limited data, as it allows for robust statistical inferences and adequately handles mixed uncertainty in the observational data. To set a meaningful prior on the mass of the system, we first established a conservative lower limit for the total mass of the close binary (A+B). This lower limit was determined by identifying the 0.01\% quantile of the posterior distribution samples obtained through MCMC analysis of EC\,95A and EC\,95B, resulting in a value of 3.7\,M$_\odot$. This approach serves as a robust prior constraint for subsequent MCMC analysis of EC\,95 (A+B+C), ensuring a meaningful estimation of the total system mass. The parameters estimated from this analysis are presented in Table \ref{T_ec95c}. Figure \ref{Orbits_ec95c} shows the orbits derived from 2,000 MCMC samples, while the posterior corner plot is provided in Figure \ref{corner_plot2} in Appendix \ref{app:1}. The total mass of the system (A+B+C) resulting from this analysis is 4.76 $^{+0.45}_{-0.36}$ M$_\odot$. 

The mass of the third component, EC\,95C, can then be determined by subtracting the mass of the close binary from the total mass of the system. To ensure accurate propagation of uncertainties and account for the non-normality of the distributions, we used the posteriors from the \orbitize fits to both the complete system and the central (A+B) sub-system, and we used  kernel density estimation (KDE) to smooth the posterior samples and calculated the difference in masses by drawing samples from these KDEs. This approach provides a robust representation of the underlying distributions and their uncertainties. The resulting estimated mass for EC\,95C is 0.26 $^{+0.53}_{-0.46}$ M$_\odot$. Our results suggest that EC\,95C has a significantly smaller mass compared to the close binary EC\,95A and EC\,95B, and is most likely a sub-solar mass star. It has a long orbital period of 172.3 $^{+13.5}_{-13.8}$ years and a fairly elliptical orbit projected on the plane of the sky. These results represent the first reliable estimates of the mass of this third component in the EC\,95 system.

\begin{table}
	\renewcommand{\arraystretch}{1.1}
\caption{Best-fit model parameters for EC\,95C using \orbitize}
\label{T_ec95c}
\centering
\small
\begin{tabular}{cccc} \hline \hline
& Parameter & Value & Units \\ 
\hline
\multicolumn{2}{l}{Orbital Parameters}&&\\
& $a$ & 52.19 $^{+2.01}_{-2.17}$ & AU \\
& $e$ & 0.75 $\pm 0.03$& \\
& $\Omega$ & 169.23$^{+7.71}_{-12.44}$ & degrees \\
& $i$ & 37.22$^{+3.15}_{-5.26}$ & degrees \\
& $\omega$ & 43.49$^{+11.03}_{-7.34}$ & degrees \\
& $T_0$ & 2458881.90 $^{+2.99}_{-3.33}$ & Julian date\\
& P & 172.30 $^{+13.45}_{-13.82}$ & years\\
& $M_{A+B+C}$ & 4.76 $^{+0.45}_{-0.36}$ & $M_{\odot}$  \\      
\hline
\end{tabular}  
	
\end{table}

\subsection{Stellar evolution models and spectral energy distribution (SED) of EC\,95}

The determination of dynamical masses in binary systems holds significant importance in refining stellar evolution models, especially for young stars, given that pre-main sequence evolution models are continuously evolving \citep[e.g.,][]{hillenbrand2004,stassun2014}. Although there have been recent developments, observational constraints for young stars of intermediate masses (2 to 8\,M$_\odot$) remain limited. Therefore, comparing the observational characteristics of EC\,95 with the predictions of theoretical models based on the dynamically determined masses presented in this study provides valuable insights.


\citet{Preibisch1999} provided initial estimates of the spectral type and bolometric luminosity corrected for extinction for EC\,95 using near-infrared spectroscopic data. According to their findings, EC\,95 was classified as a K2$\pm$2 (K0 – K4) spectral type star with a luminosity of approximately $60^{+30}_{-20}$ $L_{\odot}$, assuming a distance of 310 pc to the Serpens region. Using HR diagrams he estimated a mass of approximately $4$\,M$_{\odot}$ and an age of $0.2^{+0.2}_{-0.1}$ million years for EC\,95. Using the revised distance determined in this study of $435.71 \pm 7.43$ pc, the luminosity increases to $118.5^{+59.4}_{-39.7}$ $L_{\odot}$. \citet{Doppmann2005} conducted high-resolution near-infrared spectroscopic observations to determine the effective temperature of EC\,95. They obtain a value of $4,400^{+115}_{-57}$\,K, consistent with the spectral type proposed by \citet{Preibisch1999}. 

\begin{figure}
\includegraphics[width=0.5\textwidth]{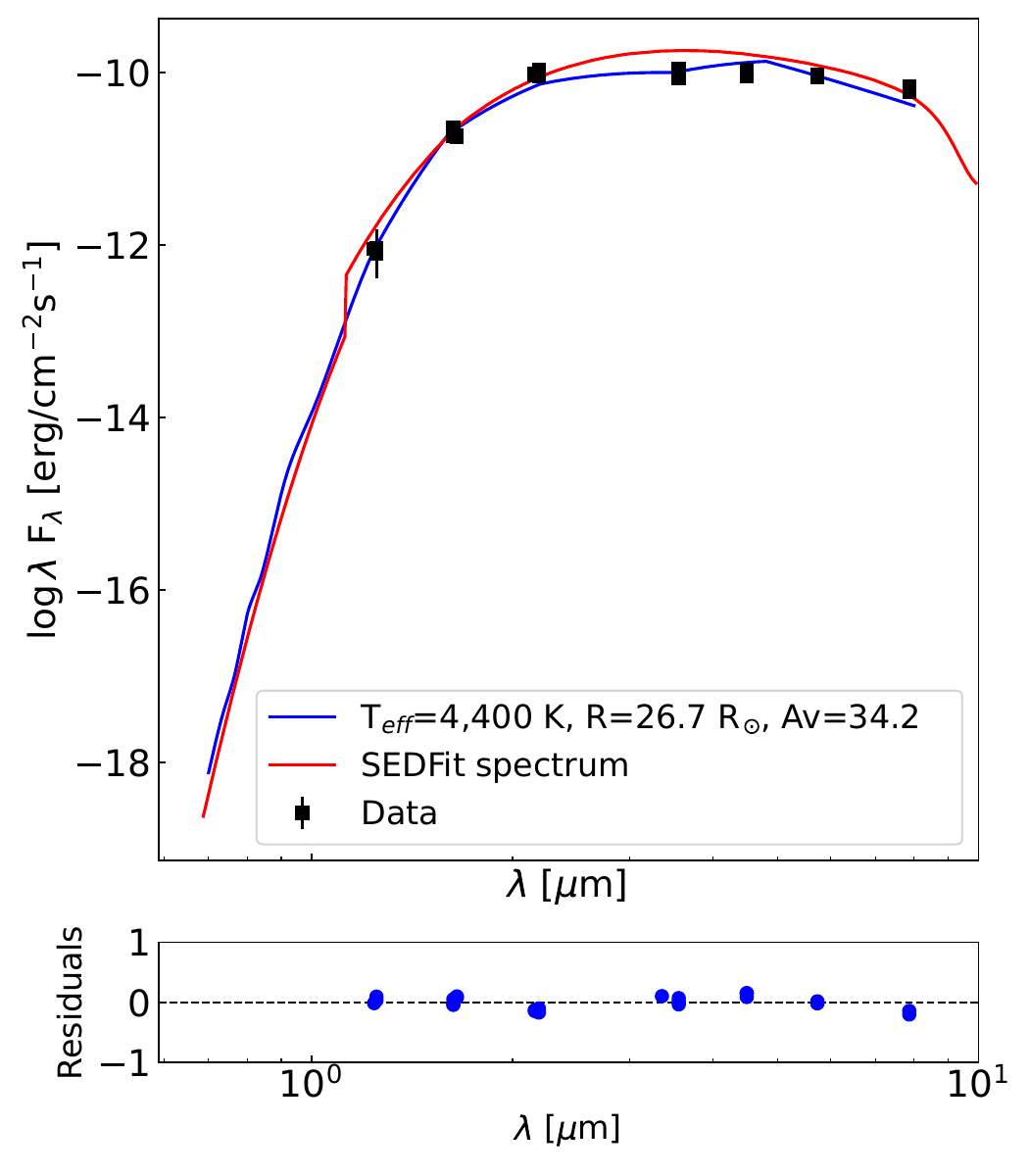}
\label{SED_ec95}
\caption{Spectral energy distribution of EC\,95. The data points (black squares) are taken from  VizieR (see Table~\ref{vizier}). The blue curve represents the SED fit with a blackbody using extinction models from \citet{Fitzpatrick2019} and \citet{rieke1989}, while the red curve corresponds to the SED fit to the spectrum using the SEDFit package. The figure also includes residual plot for the reddened blackbody model (bottom panel).}

\end{figure}

To better understand the physical properties of the EC\,95 system, we fitted its SED using a reddened blackbody model, applying the extinction models from \citet{Fitzpatrick2019} and \citet{rieke1989}. This analysis incorporated photometric data, collected through the VizieR system (see details in Table~\ref{vizier} of Appendix \ref{app:2}), covering wavelengths from 1.2 to 7.87 $\mu$m. In our fitting process, we fixed the temperature at 4,400 K and treated the stellar radius and total $V$ band extinction $A_V$ as free parameters, assuming a standard selective extinction ratio of $R_V = 3.1$. The fitting results yielded a total radius of 26.7 $\pm$ 1.1 R$_\odot$, a total luminosity of 241 $\pm$ 20 L$_\odot$, and an $A_V$ of 34.2 $\pm$ 2.5 mag, as shown by the blue curve in Figure \ref{SED_ec95}. Given that these results account for the contribution of both stars in the system, and assuming that both stars have similar luminosities due to their comparable masses, we estimate that the luminosity of each individual star is 120 $\pm$ 10 L$_\odot$ with a radius of 18.9 $\pm$ 0.8 R$_\odot$. In addition to the reddened blackbody model, we performed an SED fit based on the SEDFit package. This Python module queries VizieR\footnote{https://vizier.cds.unistra.fr} to download photometric data and fits these observational data with theoretical stellar models and extinction models (see \url{https://github.com/mkounkel/SEDFit} for details). We used the same parameters as in the reddened blackbody analysis, fixing both the temperature and stellar radius. The results from this fit show good agreement with the photometric data, as shown by the red curve in Figure \ref{SED_ec95}.

\begin{figure*}
    \centering
    \includegraphics[width=0.45\textwidth]{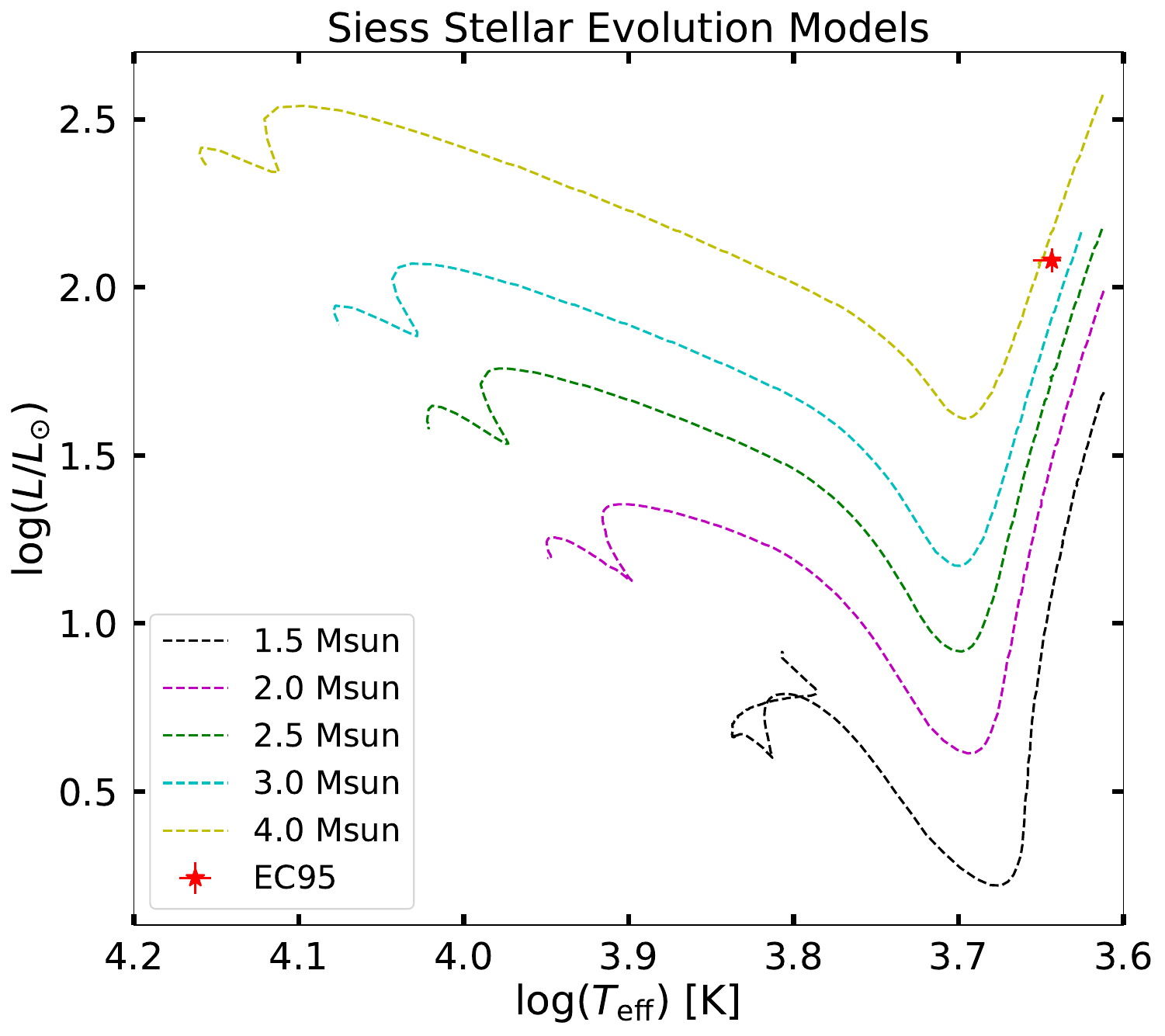}\hspace{0.2cm}
    \includegraphics[width=0.45\textwidth]{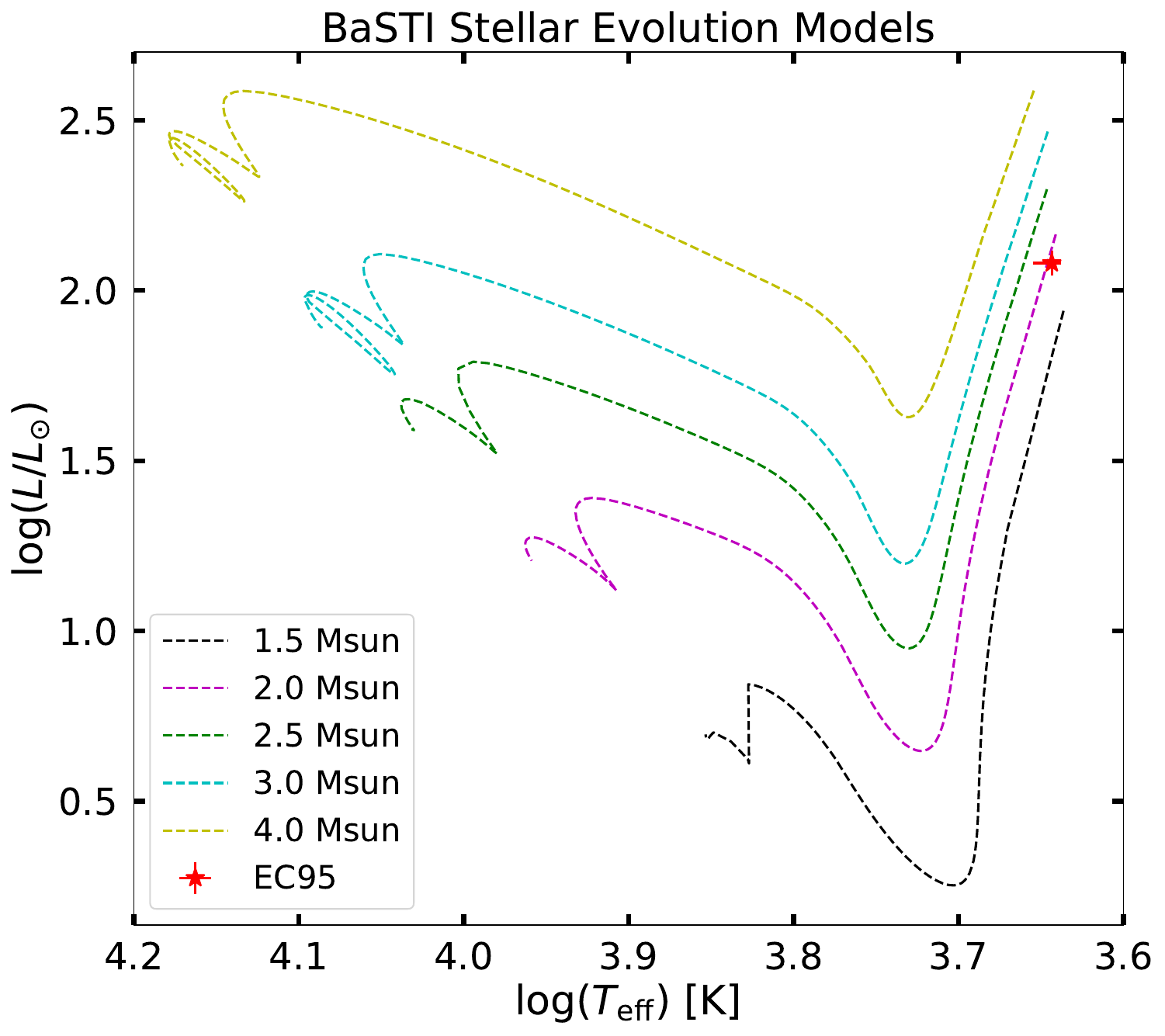}\\[-0.5ex]
    \includegraphics[width=0.45\textwidth]{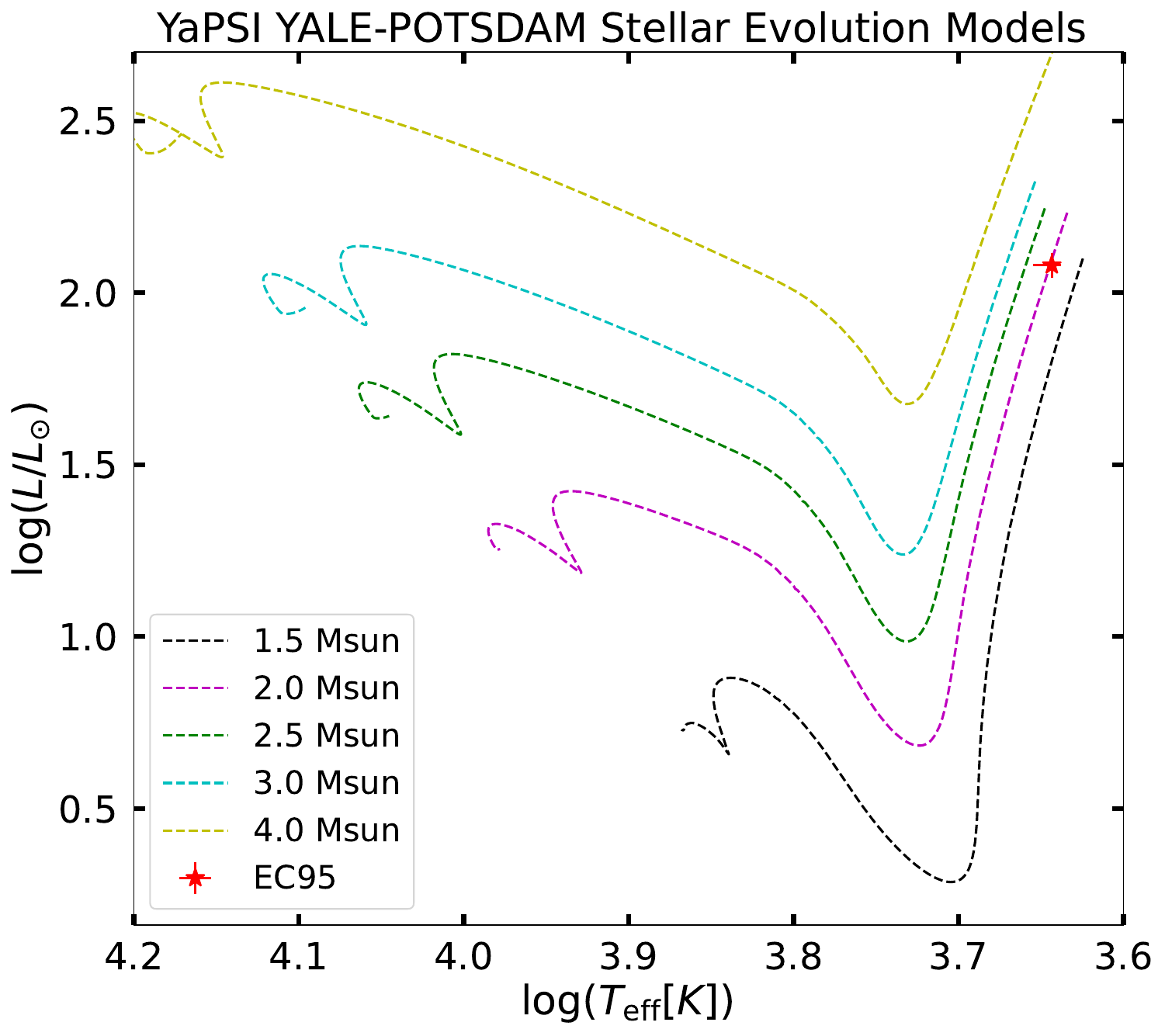}\hspace{0.2cm}
    \includegraphics[width=0.45\textwidth]{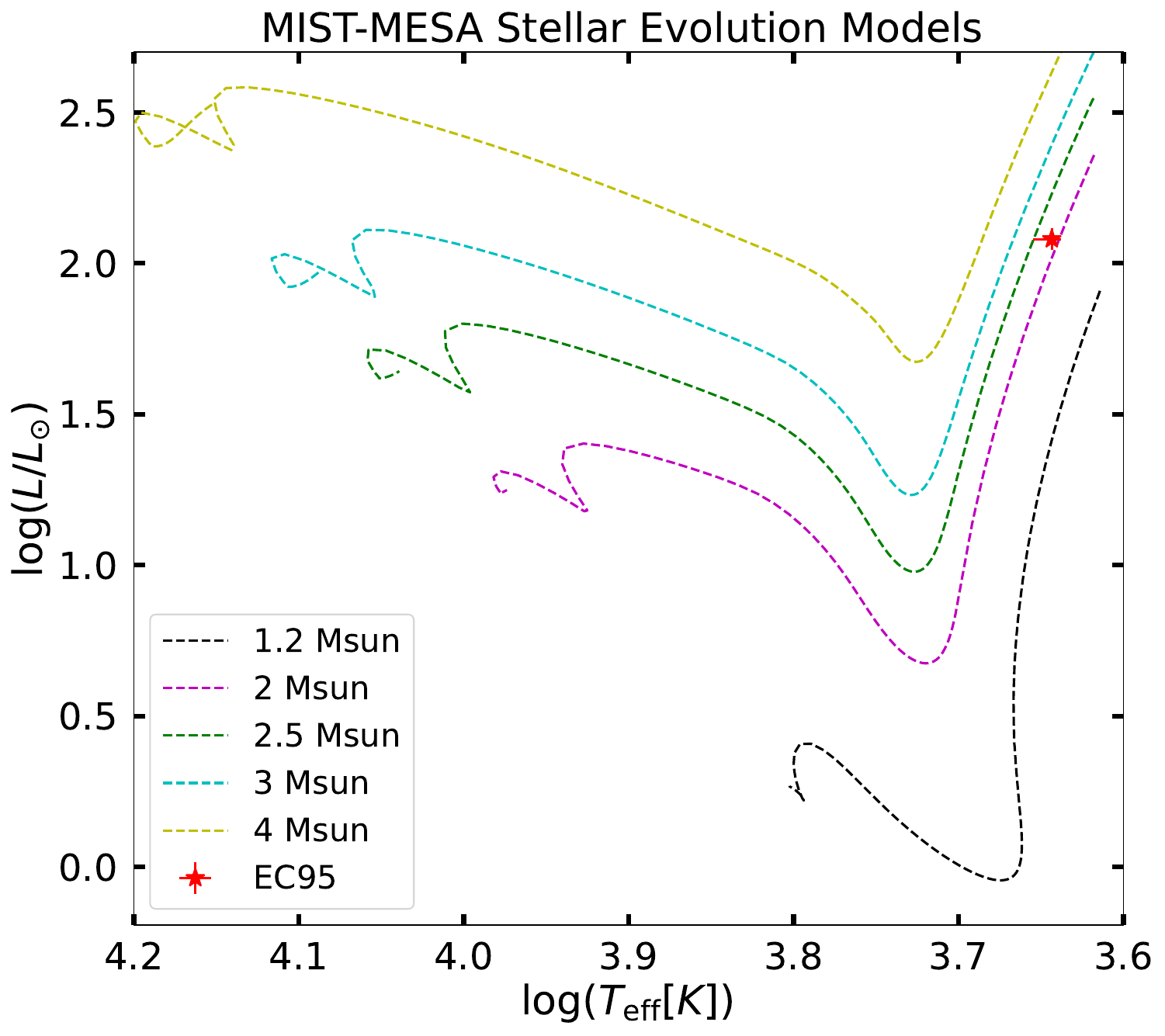}
    \caption{Location of EC\,95 on the HR diagram and evolutionary tracks based on pre-main sequence stellar evolution models. Top left: \citet{Siess2000} models. Top right: BaSTI models \citet{Pietrinferni2004,Pietrinferni2006}. Bottom right: YaPSI YALE-POTSDAM models by \citet{spada2017}. Bottom left: MIST-MESA models \citep{MESA2011,MIST2016}. These models have a metalicity $Z=0.02$, helium abundance $Y=0.2880$, mixing-length $\alpha=1.90$ and deuterium abundance $X_D=4 \times 10^{-5}$. The red star indicates the location that EC\,95 would have with  $T_{\text{eff}}$= $4,400^{+115}_{-57}$\,K and $L$ = 120 $\pm$ 10 L$_\odot$.} 
    \label{Models}
\end{figure*}

Based on the discussion above, we can place each star in EC\,95 in the HR diagram at $L = 120 \pm 10$ L$_\odot$ and $T_{\text{eff}} = 4,400^{+115}_{-57}$ K. To compare the dynamical mass measurements with theoretical predictions, we tested several pre-main sequence stellar evolution models. For each model, in Figure \ref{Models}, we show pre-main sequence evolutionary tracks for several masses. The models BaSTI \citep{Pietrinferni2004,Pietrinferni2006}, YaPSI Yale-Potsdam \citep{spada2017}, and MIST-MESA \citep{MESA2011,MIST2016} properly locate the position of each star in EC\,95 on a 2 M$_{\odot}$ track, as expected from the dynamical mass. In contrast, the models of \citet{Siess2000} would require the stars in EC\,95 to have masses between 3 and 4 M$_{\odot}$, inconsistent with the dynamical determination. We note, additionally, that the age of the stars indicated by the YaPSI and MIST-MESA models, given their position along the evolutionary track, are between 5,000 and 10,000 years, indicating that EC\,95 is a very young stellar system indeed.


\section{Discussion and conclusions} 

This work presents a comprehensive study of the astrometry of the young stellar system EC\,95 in the Serpens core region based on more than 30 individual VLBA observations collected over the last 12 years. These observations yield a distance to the system of 435.71\,$\pm$\,7.43 pc, fully consistent with the previous measurement of \citetalias{Ortiz2017}. They also confirm the hierarchical triple nature of EC\,95 and enable dynamical mass measurements of all three members of the system.

Our improved determination of the dynamical masses of the two members in the EC\,95AB compact central sub-system confirms the conclusion first presented by \citetalias{Ortiz2017} that the system is composed of two stars with similar masses of order 2 M$_\odot$. Spectroscopic observations \citep{Preibisch1999,Doppmann2005} demonstrate that EC\,95AB has an effective temperature of about 4,400 K corresponding to a spectral type of order K2. In addition, the luminosity of the system deduced from archival photometric data is $\sim$ 240 $L_\odot$, suggesting that each star has a luminosity of order 120 $L_\odot$. Comparison with several theoretical pre-main sequence stellar evolution models shows good agreement between the tracks for a 2 M$_\odot$ star and the locus of EC\,95AB in the HR diagram provided the system is extremely young ($\sim 10^4$ years). 

The description of EC\,95\,AB obtained from our analysis is somewhat different from that originally proposed by \citet{Preibisch1999} who (i) assumed that EC\,95\,AB was a single star (a fully justified assumption at the time); (ii) derived a significantly underestimated luminosity due to the shorter distance (310 pc vs.\ 435 pc) adopted for the Serpens region; and (iii) carried out a comparison with older pre-main sequence stellar evolution models which favored a higher mass. We note, indeed, that the models of \citet{Siess2000} shown in the present paper would still favor a higher mass for the EC\,95\,AB system given its location in the HR diagram (Figure \ref{Models}). As shown here, using the most recent observational results and theoretical models provides a consistent picture of EC\,95AB as a tight binary system comprised of two 2\,M$_\odot$ stars. It remains true, however, that EC\,95 appears to be a proto-Herbig Ae/Be object since a 2 M$_\odot$ star would have an early A spectral type once on the main sequence.

\begin{figure}
    \centering
    \includegraphics[width=0.45\textwidth]{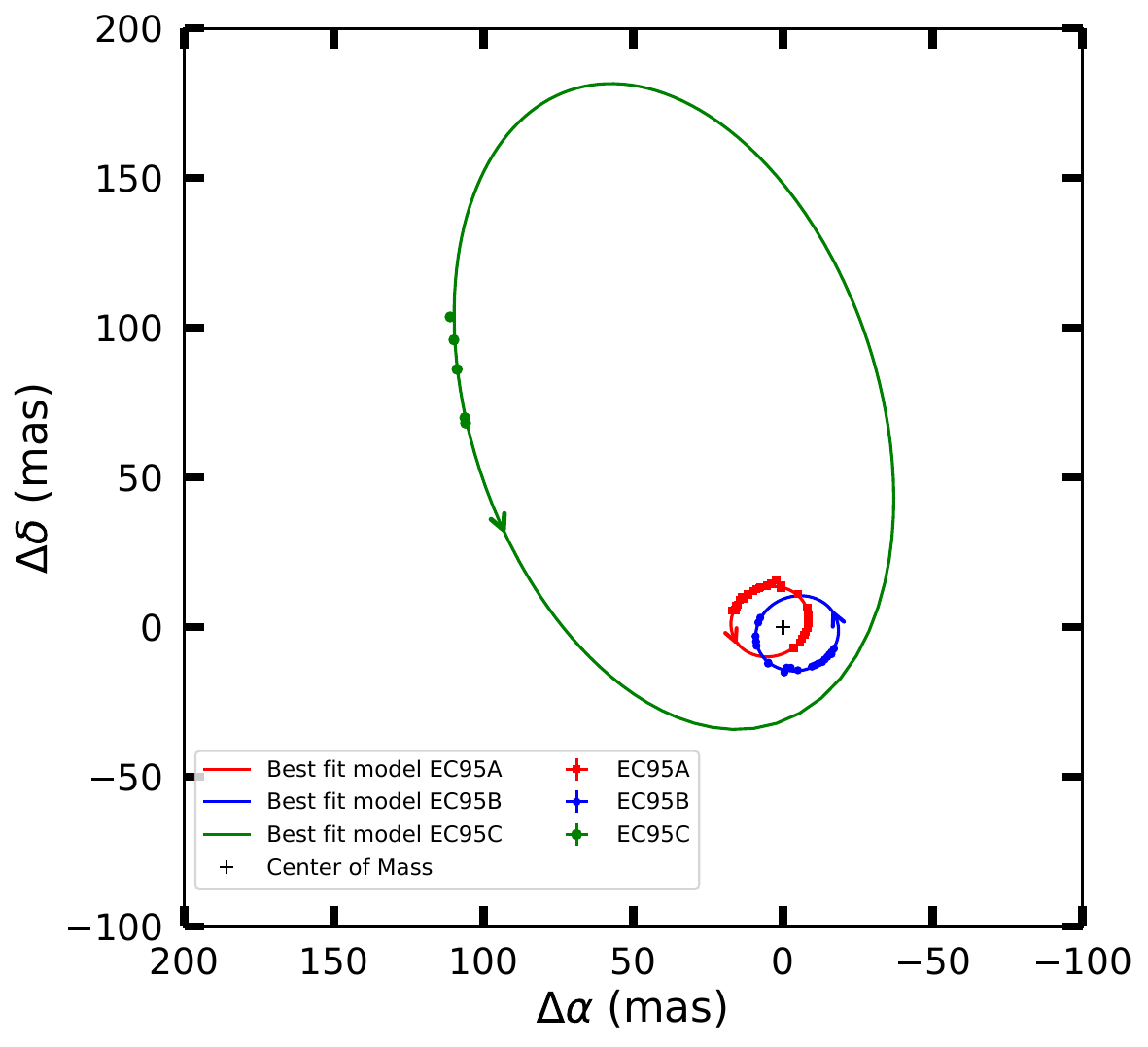}\hspace{0.2cm}
    \caption{Orbits of EC\,95A (red line), EC\,95B (blue line) and EC\,95C (green line) around the center of mass of the system (black cross). The colored squares indicate the measured positions, while the arrows show the direction of the orbits.}
    \label{fig:all_orbits}
\end{figure}

The location of EC\,95\,AB in the HR diagram is fully consistent with the derived dynamical masses of $\approx$ 2 M$_\odot$, but only provided that the system is extremely young ($\approx$ 10$^4$ yr). Before we proceed further, it is perhaps worth mentioning that this conclusion is almost inescapable because, as illustrated in Figure \ref{Models}, any star with a mass between one and several solar masses would have to be very young to find itself in that specific upper-right corner of the HR diagram. For EC\,95\,AB, such a young age is surprising because there is barely any evidence for active accretion in the system. While there is an infrared excess at 24 $\mu$m documented by Spitzer \citep{Winston2007}, near-infrared spectroscopy \citep{Doppmann2005} shows no emission lines and very low veiling. Why is such a young stellar system not accreting strongly, appearing instead as a ``naked'' protostellar system seen through substantial molecular foreground material? The tight binary nature of EC\,95\,AB could be part of the answer since tidal forces in such a system could alter the accretion processes \citep{Reipurth2014} and affect its classification. A previous example of a similar situation is provided by V773\,Tau which looks like a system of naked T Tauri stars in spite of an age more typical of a classical T Tauri system \citep{Torres2012}. Further observations will be needed to further assess the evolutionary state of EC\,95\,AB and the impact of binarity on its classification.
 
Our results are also relevant to the origin of the radio emission in EC\,95AB. Young stars with masses below about 2 M$_\odot$ can generate strong superficial magnetic fields through the dynamo effect because they possess convective interiors \citep{Feigelson1999, Palla1993}. Higher (intermediate-) mass young stars, on the other hand, are expected to be fully radiative and cannot, in principle, maintain a dynamo \citep[see, however,][]{Stello2016,Ordonez-Toro}. Thus, intermediate-mass young stars are not expected to be non-thermal radio emitters. Our results place the stars in EC\,95\,AB in a mass range where strong superficial magnetic fields are more expected than in higher-mass stars (e.g., $\sim$4 M$_\odot$). In addition, we have shown that EC\,95\,A and  EC\,95\,B likely have very large radii of order 19 R$_\odot$. This could well further favor the existence of a convective layer that could help maintain the dynamo mechanism, in a process similar to that observed in FK Comae Berenices stars \citep[e.g.][]{Hackman2013}.

Our data also enable us, for the first time, to estimate the mass of the third component in the system, EC\,95C. By calculating the total mass of the triple system and subtracting the mass of the binary system obtained with \orbitize, we estimate that EC\,95C has a mass of 0.26 $^{+0.53}_{-0.46}$ M$_\odot$. This value, considerably lower than that of the close binary stars, show that EC\,95C is the lowest mass component within the system and is most likely a T Tauri star.

A last comment is in order here. Figure \ref{fig:all_orbits} shows the orbits of all three stars in the system (A, B, and C) around the center of mass.  The inclinations of the AB orbit, on the one hand, and of the C orbit around the AB barycenter, on the other, appear similar (of order 35$^\circ$; see Tables \ref{T_A0}, \ref{mcmc} and \ref{T_ec95c}) as would be expected for a compact multiple system formed through disk fragmentation \citep{Adams1989}. Both orbits are quite eccentric, and the orientation of their semi-major axes are significantly misaligned (see the values of $\omega$ and $\Omega$ again in Tables \ref{T_A0}, \ref{mcmc} and \ref{T_ec95c}). This leads to the question of the hierarchical nature of the system (assumed for the fitting of the C orbit) and its stability. Figure \ref{fig:all_orbits} shows that, near the periastron of the C orbit, the separation between C and AB is of the same order as the separation between A and B themselves. Such a configuration could easily be unstable. Given the very limited coverage of the orbital path of EC\,95C available at the moment, there remains significant uncertainties on the orbital elements of C, so we cannot definitely conclude on the stability of the system. Resolved observations in the coming decades, both at radio wavelengths and in the infrared will be critical to examine this issue.

\section*{Acknowledgements}
{\small
J.O., L.L., G.O.L., J.M.M. and L.F.R. acknowledge the financial support of CONAHCyT, M\'exico. L.L. and S.S., acknowledge the financial support of DGAPA, UNAM (project IN112417 and IA104824, respectively). S.A.D. acknowledges the M2FINDERS project from the European Research Council (ERC) under the European Union's Horizon 2020 research and innovation programme (grant No 101018682). P.A.B.G. acknowledges financial support from the São Paulo Research Foundation (FAPESP) under grants 2020/12518-8 and 2021/11778-9. 
The Very Long Baseline Array is a facility of the National Science Foundation operated under cooperative agreement by Associated Universities, Inc. The National Radio Astronomy Observatory is a facility of the National Science Foundation operated under cooperative agreement by Associated Universities, Inc. 
}


\section*{Data Availability}
The data underlying this article were collected as part of the DYNAMO-VLBA project and can be made available upon reasonable request to the corresponding author.

\appendix
\section{}
\label{app:1}
\begin{figure*}
\includegraphics[width=\textwidth]{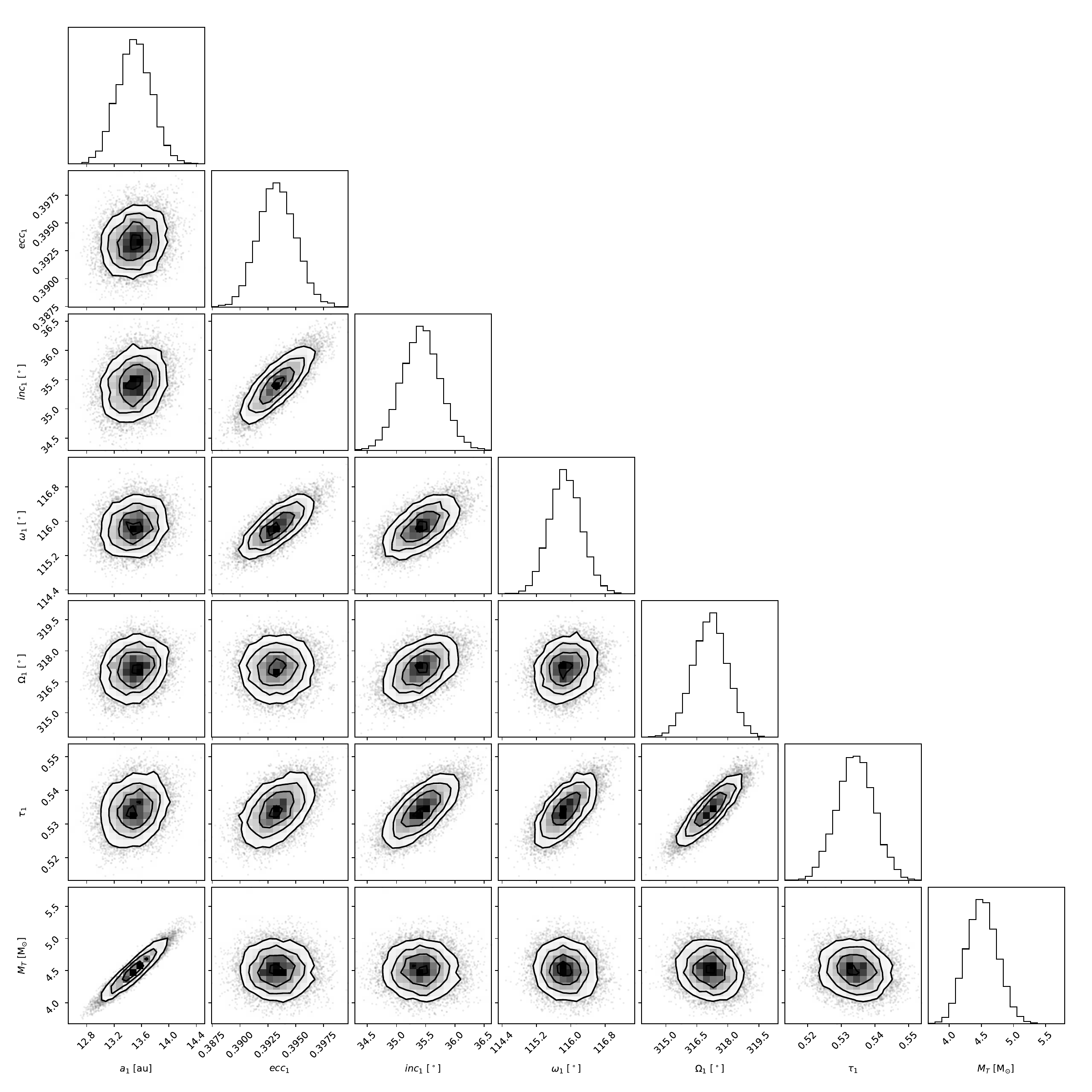}\\
\label{corner_plot}
\caption{Posterior corner plot generated from the MCMC analysis using the \orbitize package for the close binary EC\,95AB. The diagonal panels display 1D marginalized posterior distributions for each orbital parameter, including $a$, $e$, $i$, $\omega$, $\Omega$, $\tau$ and total mass. The off-diagonal panels illustrate 2D covariances between these parameters.}
\end{figure*}

\begin{figure*}
\includegraphics[width=\textwidth]{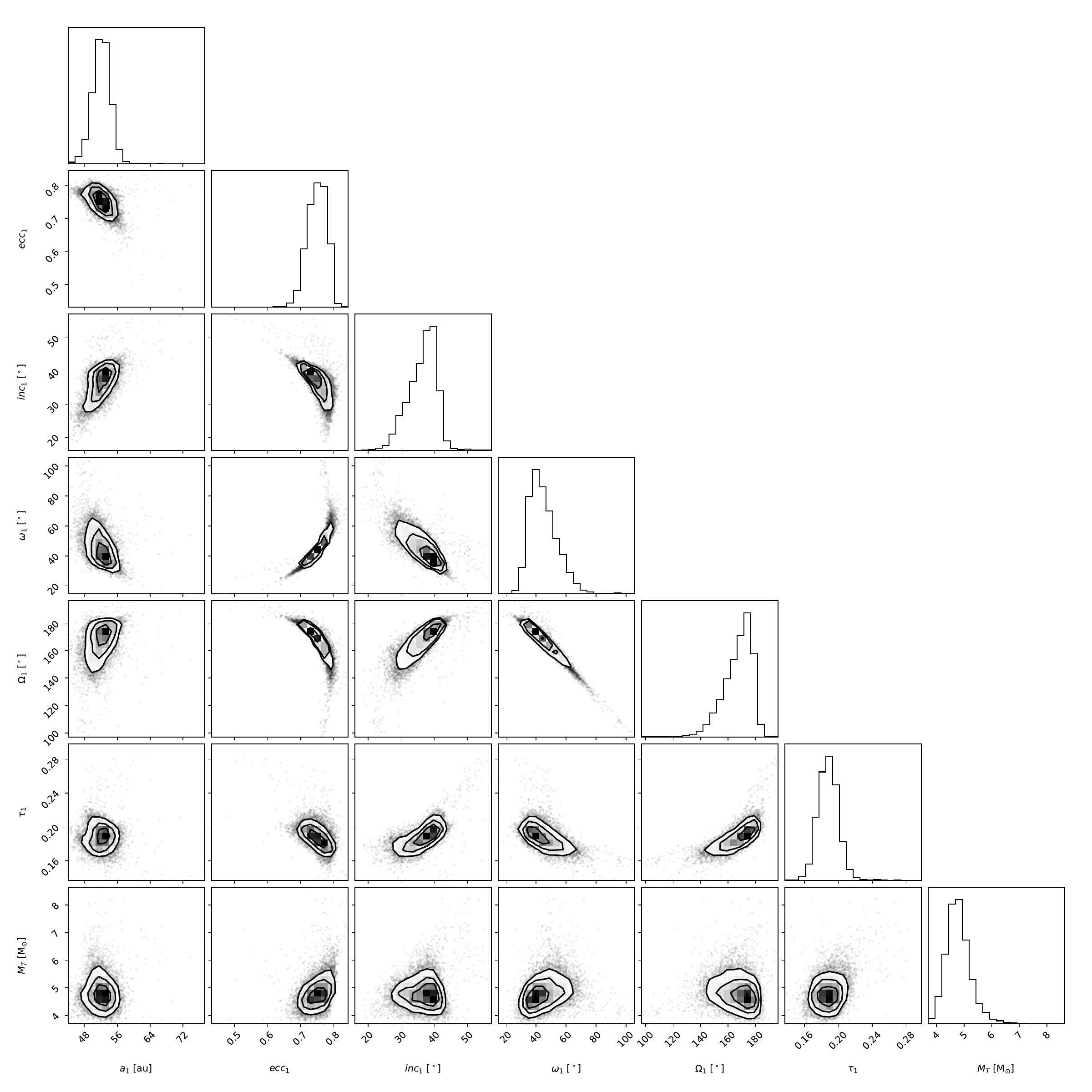}\\
\label{corner_plot2}
\caption{Posterior corner plot generated from the MCMC analysis using \orbitize for the total system mass of the triple system. The diagonal panels show marginalized posterior distributions for each orbital parameter ($a$, $e$, $i$, $\omega$, $\Omega$, $\tau$, and total mass). Off-diagonal panels display covariances between these parameters.}
\end{figure*}

In this appendix, we present the corner plots obtained from the MCMC analysis performed with the \orbitize package for the EC\,95AB and EC\,95C components of the triple system. These plots show the posterior distributions of the orbital parameters derived from the MCMC exploration. For EC\,95A, the analysis was based on 23 simultaneous observations of both components, while for EC\,95C, we included five detections (4 VLBA + 1 VLT), where the position of EC\,95C was measured relative to the barycenter of EC\,95AB, which was determined at each epoch from the astrometric results of the VLBA measurements using MPFIT. The corner plots in Figures \ref{corner_plot} and \ref{corner_plot2} correspond to EC\,95AB and EC\,95C, respectively. 

\section{Photometric Data Table for SED Analysis}\label{app:2}

The VizieR library compiles a series of modern Astronomy catalogs accessible from its webpage\footnote{\url{https://vizier.cds.unistra.fr/}}. We have used this system to locate the photometric data related to the EC\,95 system. 

All data were carefully reviewed to ensure that they correspond to our target source, avoiding duplications and mixes of nearby sources (restricting our search to a radius of $2''$ around the target position). We notice that the nearest known infrared source to EC\,95 system is EC\,92, about $5''$ to the north of the system \citep[e.g.,][\citetalias{Ortiz2017}]{Eiroa1992}, thus our search will discard data associated with this source.  We also evaluated the data in the mid-/far-infrared, such as those reported at 23 $\mu$m, but excluded them from our analysis because they are contaminated by circumstellar material.

\begin{table}
\caption{VizieR EC\,95 Photometric Data.}
\footnotesize
\begin{center}
 \setlength{\tabcolsep}{2.8pt}
 \renewcommand{\arraystretch}{1.05}
\begin{tabular}{lcccc}
\hline
\textbf{$\lambda$ ($\mu m$)} & {Flux (Jy)} & {Error flux (Jy)} & {SED Filter} & {ID Source} \\
\hline
1.25 & 0.000336 & 0.000220 & Johnson:J & \citet{Giovannetti1998} \\
1.63 & 0.0125 & 0.0001 & Johnson:H &  \\
2.19 & 0.0785 & 0.0007 & Johnson:K &  \\
\hline
1.65 & 0.0100 & 0.0004 & 2MASS:H & \citet{Evans2003} \\
2.16 & 0.0686 & 0.0022 & 2MASS:Ks &  \\
3.55 & 0.131 & 0.006 & Spitzer/IRAC:3.6 & \\
4.49 & 0.160 & 0.008 & Spitzer/IRAC:4.5 & \\
5.73 & 0.181 & 0.009 & Spitzer/IRAC:5.8 & \\
7.87 & 0.180 & 0.010 & Spitzer/IRAC:8.0 &  \\
\hline
1.24 & 0.000361 & \nodata & 2MASS:J & \citet{Giardino2007} \\
1.65 & 0.0105 & \nodata & 2MASS:H & \\
2.16 & 0.0675 & \nodata & 2MASS:Ks & \\
\hline
1.24 & 0.000377 & \nodata & 2MASS:J & \citet{Winston2007} \\
1.65 & 0.0103 & 0.0004 & 2MASS:H & \\
2.16 & 0.0694 & 0.0022 & 2MASS:Ks   \\
\hline
3.55 & 0.116 & 0.001 & Spitzer/IRAC:3.6 & \citet{Gutermuth2009} \\
4.49 & 0.145 & 0.001 & Spitzer/IRAC:4.5 &  \\
5.73 & 0.169 & 0.002 & Spitzer/IRAC:5.8 &  \\
7.87 & 0.210 & 0.002 & Spitzer/IRAC:8.0 &  \\
\hline
4.49 & 0.160 & 0.016 & Spitzer/IRAC:4.5 & \citet{Dunham2015} \\
5.73 & 0.180 & 0.018 & Spitzer/IRAC:5.8 &  \\
7.87 & 0.180 & 0.018 & Spitzer/IRAC:8.0 &  \\
\hline
3.55 & 0.103 & 0.008 & Spitzer/IRAC:3.6 & \citet{Getman2017} \\
4.49 & 0.139 & 0.006 & Spitzer/IRAC:4.5 &  \\
5.73 & 0.171 & 0.006 & Spitzer/IRAC:5.8 &  \\
\hline
\end{tabular}
\end{center}
\label{vizier}
\end{table}

\end{document}